\documentclass[11pt]{article}

\usepackage[preprint]{acl}

\usepackage{times}
\usepackage{latexsym}
\usepackage[T1]{fontenc}
\usepackage[utf8]{inputenc}
\usepackage{microtype}
\usepackage{inconsolata}

\usepackage{url}
\usepackage{booktabs}
\usepackage{titletoc}
\usepackage{multicol}
\usepackage{multirow}
\usepackage{enumitem}
\usepackage{amsmath}
\usepackage[dvipsnames,svgnames]{xcolor}
\usepackage{graphicx}
\usepackage{subcaption}
\usepackage{seqsplit}
\usepackage{textcomp}
\usepackage{dashrule}
\usepackage{tcolorbox}
\usepackage{dblfloatfix}
\usepackage{cuted}
\usepackage{placeins}
\tcbuselibrary{breakable, skins, raster, listings}

\usepackage{booktabs}
\usepackage{makecell}
\usepackage{pifont}

\definecolor{mainborder}{HTML}{2E3440}
\definecolor{contextbg}{HTML}{E6F4EA}
\definecolor{contextdark}{HTML}{2D6A4F}
\definecolor{goal1bg}{HTML}{E8F0FE}
\definecolor{goal1border}{HTML}{C5D0E6}
\definecolor{goal1dark}{HTML}{1A56DB}
\definecolor{goal2bg}{HTML}{FFF4E5}
\definecolor{goal2border}{HTML}{E6D6C5}
\definecolor{goal2dark}{HTML}{B45309}

\usepackage[dvipsnames]{xcolor}
\usepackage{pifont}

\newcommand{\cmark}{\textcolor{ForestGreen}{\ding{51}}}
\newcommand{\xmark}{\textcolor{BrickRed}{\ding{55}}}

\title{GACC-Bench: Benchmarking \textbf{G}oal-\textbf{A}daptive \textbf{C}ontext \textbf{C}ompression\\Ability across Large Language Models}

\author{
  \textbf{Weiting Liu\textsuperscript{1,*,\ddag}},
  \textbf{Zhiyu Pan\textsuperscript{2,3,*}},
  \textbf{Shuhong Wang\textsuperscript{1,*}},
  \textbf{Zhenquan Yao\textsuperscript{4,*}},
\\
  \textbf{Jungang Li\textsuperscript{5}},
  \textbf{Li Zhao\textsuperscript{3}},
  \textbf{Jian Jiang\textsuperscript{6}},
  \textbf{Jianfeng Feng\textsuperscript{1}},
  \textbf{Ai Han\textsuperscript{3,\dag}},
  \textbf{Wenlian Lu\textsuperscript{1,\dag}}
\\
\\
  \textsuperscript{1}Fudan University,
  \textsuperscript{2}Huazhong University of Science and Technology,
  \textsuperscript{3}JD.com,
\\
  \textsuperscript{4}Harbin Institute of Technology,
  \textsuperscript{5}The Hong Kong University of Science and Technology (Guangzhou),
\\
  \textsuperscript{6}Changchun Institute of Optics, Fine Mechanics and Physics, Chinese Academy of Sciences
\\
  \textsuperscript{*}Equal contribution.
  \textsuperscript{\ddag}Project leader.
  \textsuperscript{\dag}Corresponding authors.
\\
  \texttt{hanai5@jd.com, wenlian@fudan.edu.cn}
}

\begin{document}
\maketitle
\begin{abstract}
Context compression in long-running applications of large language models (LLMs) must preserve goal-relevant information, often before specific downstream questions are known. We introduce GACC-Bench, a benchmark for \textbf{G}oal-\textbf{A}daptive \textbf{C}ontext \textbf{C}ompression that evaluates how reliably LLMs adapt information retention to specified high-level goals. For each source context, we derive one or more high-level goals by grouping source-grounded questions according to shared downstream objectives. The benchmark contains 500 goals and 2,571 questions over 281 document and agent-trace contexts from seven collections spanning diverse domains, with all instances reviewed by human experts. The compressor receives a source context, a goal, and a context-length budget, with the associated evaluation questions withheld. A fixed reader answers these questions individually using the same compressed context. Our evaluation of ten LLM compressors reveals substantial gaps to full-context performance even for the best-performing model, particularly in supporting correct answers to every question associated with a goal. Compared with goal-agnostic compression, target-goal conditioning consistently improves accuracy, whereas mismatched-goal conditioning reduces it. GACC-Bench is available at \url{https://github.com/weiting24/GACC-Bench}.
\end{abstract}

\section{Introduction}
\label{sec:intro}
Large language models (LLMs)~\citep{singh2025openai,yang2025qwen3,xu2026deepseek} are increasingly deployed for long-horizon tasks~\citep{yao2023react,yang2024sweagent}, where accumulated instructions and intermediate findings can exceed the available context budget. Coding assistants such as Codex CLI and Claude Code already use context compaction in long sessions. Claude Code also allows users to specify the focus
of compaction. At compression time, users may know their next objective even when the specific questions needed to pursue it are unknown. We define a \textbf{goal} as a high-level description of the intended future use of a source context. Goal-adaptive compression aims to preserve information for multiple subsequent questions associated with that goal.

\begin{figure*}
    \centering
    \includegraphics[width=1.0\linewidth]{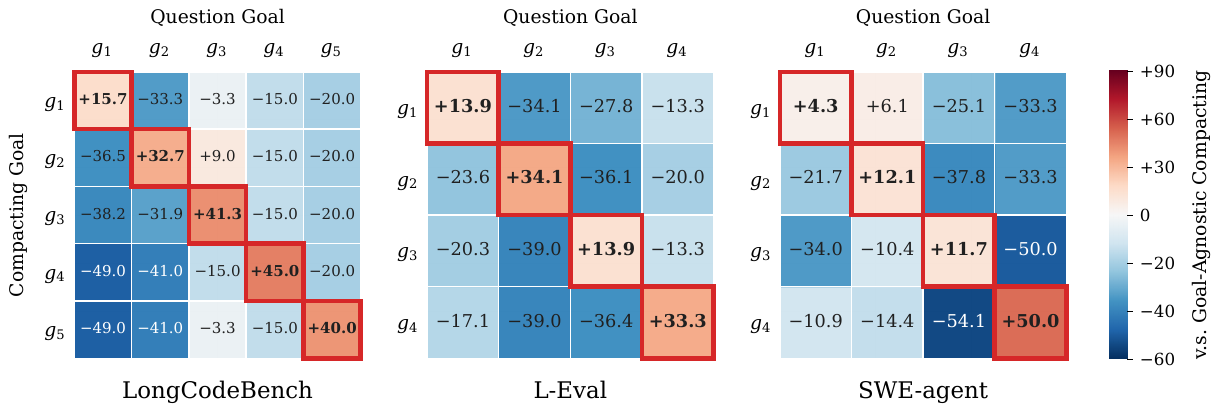}
    \vspace{-18pt}
    \caption{\textbf{Goal-specific gains from goal-guided compression.} Results are based on 20 contexts from each of three sources: LongCodeBench, L-Eval, and agent trajectories collected from SWE-Bench Verified tasks. DeepSeek-V4-Pro serves as both the compressor and reader. Rows indicate compression goals, while columns indicate the goals associated with the evaluation questions. Each cell reports the change in question-level accuracy relative to goal-agnostic compression, aggregated across contexts and expressed in percentage points. Red outlines mark matched goals. All matched-goal cells show gains, whereas 42 of the 44 mismatched-goal cells show losses.}
    \label{fig:motivation_exp}
\end{figure*}

Prior work has studied query-aware~\citep{jiang2024longllmlingua,xu2024recomp}, query-agnostic~\citep{pan2024llmlingua}, and task-aware~\citep{corallo2026cachenotes} compression. However, it remains unclear how reliably a compressed context supports multiple withheld questions for a context-specific goal, and whether gains depend on which goal guides compression. In a motivating experiment, we use DeepSeek-V4-Pro~\citep{xu2026deepseek} to compress each context separately for each associated goal and evaluate each compressed context on questions associated with all goals for that context. Figure~\ref{fig:motivation_exp} reports changes in question-level accuracy relative to goal-agnostic compression: matched-goal compression consistently improves accuracy, whereas most mismatched-goal conditions reduce it. These results indicate that \textit{the gains from goal-guided compression are goal-specific}.

Motivated by these findings, we introduce GACC-Bench to evaluate \textbf{G}oal-\textbf{A}daptive \textbf{C}ontext \textbf{C}ompression in LLMs. We generate source-grounded questions with reference answers and supporting evidence, group them by shared downstream objectives, and formulate a high-level goal for each group. After quality control, each source context is paired with one or more goals and their associated questions. Figure~\ref{fig:benchmark_example} illustrates how different goals for the same context prioritize different information. At evaluation time, the compressor receives the source context, a goal, and a token budget, without access to the associated evaluation questions. A fixed reader answers these questions individually using the same compressed context, and a fixed judge evaluates answer correctness against the original source context and reference answers.

Our contributions are threefold. First, we introduce
GACC-Bench, to the best of our knowledge, the \textbf{first benchmark} for goal-adaptive context compression. It contains 500 goals and 2,571 questions over 281 document and agent-trace contexts from seven collections spanning diverse domains, with all instances reviewed by human experts. Second, we design a compressor--reader--judge framework to evaluate goal-adaptive compression under different retention budgets, with the reader and judge held fixed. Third, we evaluate ten LLM compressors and conduct goal ablations across six models. Our experiments show that target-goal conditioning improves accuracy over goal-agnostic compression, yet even the best-performing compressor falls substantially short of full-context performance in supporting correct answers to every question associated with a goal.

\begin{figure*}[ht]
\centering
\begin{tcolorbox}[
    enhanced,
    colback=white,
    colframe=mainborder,
    boxrule=1pt,
    arc=4pt,
    boxsep=2pt,
    left=2pt, right=2pt, top=2pt, bottom=2pt,
    title={\sffamily\bfseries\small Example Goals with the Corresponding Questions in GACC-Bench},
    coltitle=mainborder,
    colbacktitle=white,
    fonttitle=\bfseries,
    titlerule=1pt,
    fontupper=\small 
]

\textbf{\textsf{\textcolor{contextdark}{CONTEXT:}}}
\vspace{-4pt}
\begin{tcolorbox}[
    enhanced,
    colback=contextbg,
    colframe=contextbg,
    boxrule=0pt,
    arc=2pt,
    left=2pt, right=2pt, top=2pt, bottom=2pt
]
\rmfamily
\tiny Is the government exhausting its powers? An empirical examination of eminent domain exercises in New York
City pre- and post-Kelo\\
Abstract:\\
A controversial U.S. Supreme Court decision in Kelo v. City of New London (2005) which did not limit the use of state’s eminent domain powers, led to an unprecedented legislative reaction by almost all 50 states. Of all, New York State stands out as one of the single states not to respond with a legislative amendment. In this study, I ask whether the state’s predation was greater in the years following these legal and political developments, in light of the freedom which was granted to local politicians by both the Supreme Court and the state’s legislators. The article hypothesizes that contrary to common perceptions, judicial decisions impact local government actions even when no limits on the use of powers are being posed. I use rigorous statistics and scrupulously deﬁned data to expand scholarly understanding of the aftermath of the judicial decision in Kelo. The main ﬁnding is that the decision has …
\end{tcolorbox}

\vspace{-3mm}

\begin{tcbitemize}[
    raster columns=2,      
    raster column skip=2mm, 
    raster equal height,    
    empty,
    boxrule=0pt,
    left=0pt, right=0pt, top=0pt, bottom=0pt
]

\tcbitem
    \textbf{\sffamily\textcolor{goal1dark}{GOAL 1:}} \vspace{2.5mm} \\ \tiny \textbf{\textcolor{goal1dark}{Understand how New York City's post-Kelo eminent domain practices compared to national trends and state-level legislative and judicial responses, and what the study's empirical findings reveal about the hypothesis that city officials would become more aggressive in their takings absent constitutional impediments.}}

    \vspace{+6pt}
    
    \begin{tcolorbox}[
        enhanced,
        colback=goal1bg,
        colframe=goal1border,
        boxrule=0.5pt,
        arc=2pt,
        boxsep=2pt,
        left=2pt, right=2pt, top=2pt, bottom=2pt
    ]
        \textbf{\textcolor{goal1dark}{Question 1:}}\\
        According to the source, how did the growth patterns observed for New York compare to those across the country, and what specific national estimate did Berliner provide for the year following the Kelo decision that served as the basis for that comparison?\\
        \textbf{\textcolor{goal1dark}{Standard Answer:}}\\
        The comparison of New York growth patterns to national growth patterns: The growth patterns observed for New York were not dramatically different than ...
        
    \end{tcolorbox}
    
    
    \begin{tcolorbox}[
        enhanced,
        colback=goal1bg,
        colframe=goal1border,
        boxrule=0.5pt,
        arc=2pt,
        boxsep=2pt,
        left=2pt, right=2pt, top=2pt, bottom=2pt
    ]
        \textbf{\textcolor{goal1dark}{Question 2:}}\\
        According to the source, what was the general post-Kelo pattern of state legislative reform across the United States, how did New York State's legislative response compare to that pattern, and what specific action by New York State's highest court made the state an exception to the general post-Kelo pattern of stronger state judicial scrutiny of public use issues?\\
        \textbf{\textcolor{goal1dark}{Standard Answer:}}\\
        The general post-Kelo pattern of state legislative reform across the United States: Almost all states enacted post-Kelo reforms to their eminent domain state laws ...
    \end{tcolorbox}

\tcbitem
    
    \textbf{\sffamily\textcolor{goal2dark}{\footnotesize GOAL 2:}} \\ \tiny \textbf{\textcolor{goal2dark}{Understand how the study's empirical findings characterize the change in New York City's eminent domain practices for urban renewal projects after the 2005 Kelo decision, including the statistical and case-specific evidence for the reported decline.}}
    
    \vspace{-0.2mm}
    
    \begin{tcolorbox}[
        enhanced,
        colback=goal2bg,
        colframe=goal2border,
        boxrule=0.5pt,
        arc=2pt,
        boxsep=2pt,
        left=2pt, right=2pt, top=2pt, bottom=2pt
    ]
        \textbf{\textcolor{goal2dark}{Question 1:}}\\
        For the simple logistic regression that used only the year as an independent predictor to assess the trend in the share of urban renewal projects, what were the stated coefficient estimate, p-value, and the substantive interpretation for the pre-2005 period (1991–2005), and what were the corresponding coefficient estimate, p-value, and substantive interpretation for the post-2005 period?\\
        \textbf{\textcolor{goal2dark}{Standard Answer:}}\\
        coefficient estimate for pre-2005 period: -0.04;\\
        p-value for pre-2005 period: 0.45; \\
        substantive interpretation for pre-2005 period...
    \end{tcolorbox}
    
    \begin{tcolorbox}[
        enhanced,
        colback=goal2bg,
        colframe=goal2border,
        boxrule=0.5pt,
        arc=2pt,
        boxsep=2pt,
        left=2pt, right=2pt, top=2pt, bottom=2pt
    ]
        \textbf{\textcolor{goal2dark}{Question 2:}}\\
        For the Goldstein (Atlantic Yards) project and the Kaur (Columbia University) project, identify: (a) which entity took the private property in each case, (b) the stated purpose of the taking for each project, (c) the year each project was first announced, and (d) the date each project's land was formally taken.\\
        \textbf{\textcolor{goal2dark}{Standard Answer:}}\\
        Entity that took private property for Goldstein project: Empire State Development Cooperation (ESDC)...\\
        Entity that took private property for Kaur project: ESDC Stated purpose for ...
    \end{tcolorbox}

\end{tcbitemize}

\end{tcolorbox}
\vspace{-2mm}
\caption{\textbf{An illustrative example from GACC-Bench.} The context is paired with multiple distinct goals, each formulated from a group of questions. Goal 1 requires retaining information about post-\textit{Kelo} national legislative trends and judicial responses, whereas Goal 2 necessitates preserving specific statistical evidence and detailed project entities.}
\vspace{-4pt}
\label{fig:benchmark_example}
\end{figure*}

\section{Related Works}
\label{sec:relatedwork}

\begin{table*}[t]
\centering
\footnotesize
\small
\caption{Positioning GACC-Bench against existing long-context and context-compression benchmarks.}
\label{tab:compression_protocol_dimensions}
\setlength{\tabcolsep}{3pt}
\renewcommand{\arraystretch}{1.05}
\begin{tabular}{@{}lccccccc@{}}
\toprule
\textbf{Benchmark}
& \makecell{\textbf{Compressor}\\\textbf{input}}
& \makecell{\textbf{Reader}\\\textbf{input}}
& \makecell{\textbf{Compression}\\\textbf{supported}}
& \makecell{\textbf{Future task}\\\textbf{withheld}}
& \makecell{\textbf{Long}\\\textbf{context}}
& \makecell{\textbf{Agent}\\\textbf{traces}}
& \makecell{\textbf{Token}\\\textbf{budget}} \\
\midrule

LongBench-v2~\citep{bai2025longbench}
& --
& Question
& \xmark
& \xmark
& \cmark
& \xmark
& \xmark \\

HELMET~\citep{yen2025helmet}
& --
& Question
& \xmark
& \xmark
& \cmark
& \xmark
& \xmark \\

LongMemEval~\citep{wu2024longmemeval}
& --
& Question
& \xmark
& \xmark
& \cmark
& \xmark
& \xmark \\

LongContextBench~\citep{yuan2024kv}
& Question
& Same question
& \cmark
& \xmark
& \cmark
& \xmark
& \cmark \\

SCBench~\citep{li2025scbench}
& Question
& Same question
& \cmark
& \xmark
& \cmark
& \xmark
& \cmark \\

LongMemEval-V2~\citep{wu2026longmemeval}
& Question
& Same question
& \cmark
& \xmark
& \cmark
& \cmark
& \cmark \\

\textbf{GACC-Bench (ours)}
& \textbf{Goal}
& \textbf{Question}
& \cmark
& \cmark
& \cmark
& \cmark
& \cmark \\

\bottomrule
\end{tabular}
\end{table*}

\paragraph{Context Compression with Downstream Guidance.}

    Early context-compression methods select content using self-information~\citep{li2023compressing} or perplexity-based importance~\citep{jiang2023llmlingua}. LLMLingua-2 learns task-agnostic token pruning~\citep{pan2024llmlingua}, whereas LongLLMLingua and RECOMP condition compression on the downstream question~\citep{jiang2024longllmlingua,xu2024recomp}. Query-based summarization also uses explicit information requests: QMSum pairs meetings with requests for relevant summaries~\citep{zhong2021qmsum}. For agents, ACON optimizes compression guidelines using trajectories that succeed with full context but fail after compression~\citep{kang2025acon}. CacheNotes develops a task-aware KV-cache compression method guided by task descriptions shared across examples within a task family~\citep{corallo2026cachenotes}. It generates compression-planning tokens to construct a reusable compressed cache before downstream queries are available.

\paragraph{Evaluating Information Retention.}

    Comparative studies assess downstream performance across compression methods and ratios~\citep{jha2024characterizing,zhang2025empirical}. Question-answering-based summary metrics evaluate factual consistency and information coverage~\citep{wang2020asking,durmus2020feqa,scialom2021questeval}. Memory benchmarks examine information use across extended interactions: LoCoMo and LongMemEval evaluate conversational recall and reasoning~\citep{maharana2024evaluating,wu2024longmemeval}, while MemoryAgentBench includes test-time learning and selective forgetting~\citep{hu2026evaluating}. SUMER retrieves from uncompressed histories to answer questions~\citep{zheng2025goal}, allowing information absent from the current compressed context to be recovered.

As summarized in Table~\ref{tab:compression_protocol_dimensions}, GACC-Bench provides a benchmark for compression guided by context-specific high-level goals across document and agent-trace contexts. It uses multiple held-out questions per goal and comparisons across goals for the same source to assess goal-specific information retention.



\section{Methodology}
\label{sec:methodology}

\begin{figure*}[t]
    \centering
    \includegraphics[width=\linewidth]{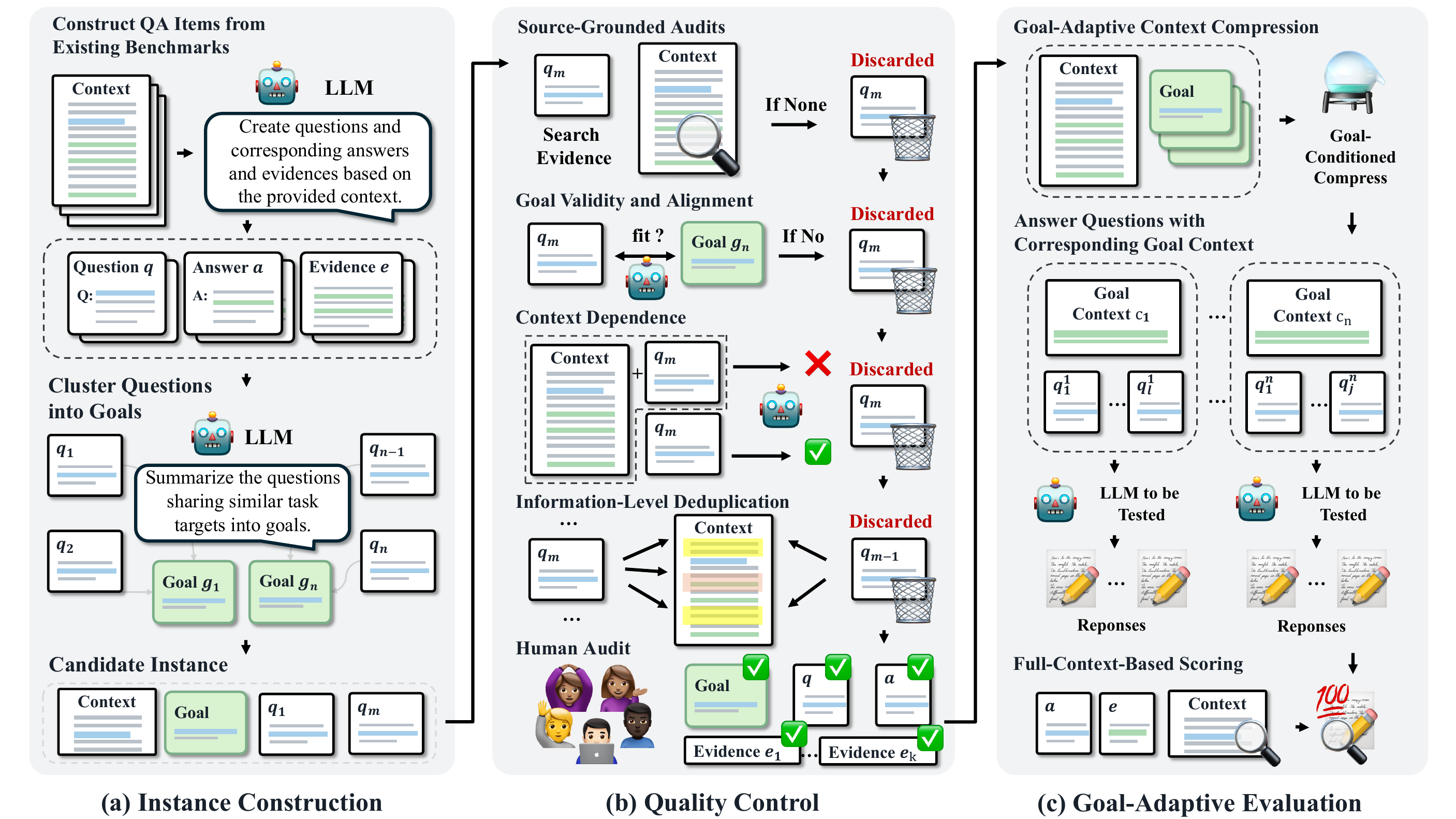}
    \caption{Instance construction, quality control, and compression evaluation of the GACC-Bench.}
    \label{fig:benchmark_framework}
\end{figure*}

We construct GACC-Bench by generating source-grounded questions, grouping them by shared information needs, and synthesizing a high-level goal for each group. Quality control checks question redundancy, within-group coherence, goal--question alignment, and cross-goal distinctiveness. Figure~\ref{fig:benchmark_framework} summarizes the three stages: instance construction, quality control, and compression evaluation.

\subsection{Instance Construction}
\label{sec:instance-construction}

Given a source context $c$ drawn from natural-language documents, code repositories, or interaction histories, we generate questions grounded in the source. A future goal may depend on a stated fact, a relation among facts, or a condition that changes how those facts should be used. To test whether compression preserves these forms of information, we use seven strategies for generating questions.

\emph{\textbf{Information retrieval}} asks for a fact explicitly stated in the source, such as a reported value or decision. \emph{\textbf{Multi-step derivation}} requires combining facts stated in the source through successive logical or numerical steps rather than recalling an explicit conclusion. \emph{\textbf{Cross-entity comparison}} compares entities on the same attributes while preserving which values belong to which entity, period, or scope. \emph{\textbf{Whole decomposition}} reconstructs a source-defined whole from its components and, where supported, derives a share or residual. \emph{\textbf{Rule application}} applies a stated rule, threshold, or exception to a specified case, including a hypothetical change to an explicit input. \emph{\textbf{Bounded trend extrapolation}} uses a stated rate, recurrence, or sufficiently specified series to derive a limited next value or range. \emph{\textbf{Conflict resolution}} identifies competing or revised claims, retains their provenance and scope, and determines which claim governs only when the source establishes an authority or temporal ordering. Each question has a reference answer and supporting evidence. A strategy is used only when it is supported by the source. The generation prompts are provided in the appendix~\ref{appendix:prompt_question_generation}.

Before question generation, candidate facts and their locations in $c$ are identified. Each question is generated with a reference answer and a record linking its requested answer components to the relevant facts and source passages. During quality control, these recorded links are used to assess source support and whether the selected evidence is sufficient to answer the question. Using the source context and generated questions, an LLM groups questions that serve a shared future task and synthesizes a goal $g$ for each group. Questions in a group may require different information but should jointly support that task. The goal describes what the retained information should enable, without listing questions or revealing answers. The resulting instance is $z=(c,g,\mathcal{Q}_g)$, where $\mathcal{Q}_g=\{q_1,\ldots,q_{n_g}\}$ contains the associated evaluation questions.

\subsection{Quality Control}
\label{sec:quality-control}

We screen instances for source validity, context dependence, question coherence, and goal alignment before compression evaluation. Checks remove invalid questions or goal groups; unrelated compound questions are separated only when their answer requirements can be assigned to coherent new questions.

\noindent\textbf{Source-Grounded Audits.}
A question is retained only if it is unambiguous, its factual premises are supported by the source, and the essential claims in its generated reference answer are correct. The recorded facts and source passages are used to check whether each requested answer component has sufficient support. To assess completeness independently of the generated answer, an LLM identifies all requested items from the question and source, including alternative correct answers when the question permits them. These requirements are audited before the final reference answer is constructed. We also reject questions whose requested relations or conditions lack source support or do not follow their assigned generation strategy. Questions about incidental details that do not serve a substantive downstream task, such as command counts or line numbers in SWE-agent histories, are removed.

\noindent\textbf{Goal Validity and Alignment.}
Each synthesized goal must describe a coherent future task grounded in the source and supported by its assigned questions. Topic labels, restatements of questions, and goals that reveal answers are rejected. Within each group, questions must fit the goal, and their answer requirements must concern information worth retaining for that task. Three models assess each requirement; any valid negative judgment removes the question. When a context has multiple goals, an LLM must select the assigned goal for each question and an assigned question for each goal from alternatives in the same context. A failed selection removes the tested question or goal, respectively. A goal is also removed if it covers information requested by a question from another group, judged from that question and its reference answer without the source. Groups with fewer than four remaining questions are discarded.

\noindent\textbf{Context Dependence.}
Context dependence is checked in both directions: the full source must enable a correct answer, while the question alone should not. Answers must be correct across repeated attempts with the full context. For questions combining source facts, the selected source passages must also suffice to answer the question; within this selection, at least two passages must contribute necessary information. To test answerability without the source, four models each receive only the question, without the goal or reference answer. Their responses are judged against the full source; any correct response removes the question. Questions that disclose their own answers are removed as well.

\noindent\textbf{Information-Level Deduplication.}
Question pairs within each goal group are compared at two levels. First, if all information requested by one question is covered by another, the more informative question is retained. Second, for questions that ask for several pieces of information, we compare their individual requests. Requests with the same wording are matched directly, while an LLM compares each request and its corresponding reference answer to identify duplicates expressed differently. If any request is duplicated across two questions, one entire question is removed, even if its other requests differ.


\paragraph{Expert Human Review.}

    A team of five human experts reviewed all benchmark instances following the automated quality checks. Each instance was reviewed by two experts to recheck all criteria covered by the automated checks, including whether questions were grounded in their source contexts, reference answers were supported by the source, and goals were aligned with their associated questions. The two reviewers initially disagreed on approximately 5\% of the reviewed goal groups. Disagreements between the two reviewers were discussed with a third expert. Instances with identified quality issues were removed.

\subsection{Compression Evaluation}
\label{sec:evaluation}

GACC-Bench evaluates goal-adaptive compression using a two-stage \textit{Compressor--Reader} framework, in which the compressor condenses the source context and the reader answers questions based on the resulting compressed context.

\paragraph{Goal-conditioned context compression.}

Token counts are measured using \texttt{cl100k\_base} as a common reference tokenizer. For a given retention ratio, the absolute token budget is calculated by multiplying the source-context length by the retention ratio and taking the floor, with a minimum budget of one token. The compressor receives the source context, the goal, and the resulting absolute token budget, and produces a compressed context that preserves goal-relevant information and relations. A single compressed context is generated for each context--goal--ratio tuple and reused for all questions associated with the same goal. After generation, the output is tokenized using the same reference tokenizer. If the generated output exceeds the budget, we retain its first $L$ tokens as the compressed memory. If the output is shorter than $L$ tokens, it is retained without padding. No length-based regeneration or rewriting is applied. We record both the raw generated length and the final retained length.

\paragraph{Compressed-context question answering.}

For each question associated with a goal, the fixed reader receives the corresponding compressed context and the question, and produces a candidate answer. A fixed LLM judge evaluates the candidate answer using the original source context, the question, and the ground-truth answer. We validate the LLM judge by comparing its decisions with human expert judgments (Appendix~\ref{app:judge_validation}).

\section{Experiments}
\label{sec:exp}

\begin{table*}[t]
\caption{Goal-adaptive compression accuracy on the
GACC-Bench-full at 5\% and 10\% retention budgets.}
\label{tab:main_table}
\centering
\small
\resizebox{\linewidth}{!}{%
\begin{tabular}{l|ccccccc|cc}
\toprule
\toprule
\multirow{2}{*}{Model}
& \multicolumn{7}{c|}{Dataset-level Macro Accuracy}
& \multicolumn{2}{c}{Overall Accuracy} \\
\cmidrule(lr){2-8}
\cmidrule(lr){9-10}
& LBv2 & HELMET & L-Eval & Loong & MB & LCB & SWE
& Macro & Micro \\
\midrule
\multicolumn{10}{c}{Budget: 5\%} \\
\midrule
GPT-6-Astra      & 14.29 & 2.50 & 0.00 & 11.76 & 0.00 & 6.41 & 17.86 & \textbf{6.00} & \underline{42.43} \\
GPT-6.1-Sol      & 14.29 & 2.50 & 1.96 & 8.82  & 0.00 & 6.41 & 10.71 & \underline{5.40} & \textbf{42.86} \\
GPT-5.6-Sol      & 6.12  & 2.50 & 1.96 & 1.96  & 0.00 & 1.28 & 0.00  & 1.80 & 29.64 \\
Claude-Opus-5.5  & 6.12  & 0.00 & 0.00 & 6.86  & 0.00 & 3.85 & 3.57  & 2.80 & 28.94 \\
Gemini-3.6-flash & 2.04  & 0.00 & 0.00 & 0.00  & 0.00 & 1.28 & 0.00  & 0.40 & 20.30 \\
Qwen3.8-Max      & 6.12  & 2.50 & 1.96 & 0.98  & 0.00 & 6.41 & 10.71 & 3.00 & 30.84 \\
GLM-5.3          & 14.29 & 0.00 & 1.96 & 7.84  & 0.00 & 5.13 & 10.71 & 4.80 & 31.19 \\
GLM-5.3-flash    & 8.16  & 0.00 & 0.98 & 4.90  & 0.00 & 1.28 & 7.14  & 2.60 & 31.43 \\
DeepSeek-V4-Pro  & 8.16  & 2.50 & 0.00 & 0.00  & 0.00 & 5.13 & 3.57  & 2.00 & 28.70 \\
Kimi-K3          & 10.20 & 2.50 & 0.98 & 2.94  & 0.99 & 3.85 & 7.14  & 3.20 & 28.67 \\
\midrule
\multicolumn{10}{c}{Budget: 10\%} \\
\midrule
GPT-6-Astra      & 30.61 & 12.50 & 11.76 & 21.57 & 0.99 & 17.95 & 17.86 & \textbf{14.80} & \textbf{57.41} \\
GPT-6.1-Sol      & 30.61 & 12.50 & 9.80  & 24.51 & 1.98 & 14.10 & 21.43 & \textbf{14.80} & \underline{56.40} \\
GPT-5.6-Sol      & 14.29 & 0.00  & 5.88  & 5.88  & 1.98 & 5.13  & 3.57  & 5.20  & 43.21 \\
Claude-Opus-5.5  & 20.41 & 7.50  & 1.96  & 14.71 & 1.98 & 11.54 & 17.86 & 9.20  & 43.95 \\
Gemini-3.6-flash & 18.37 & 2.50  & 0.00  & 1.96  & 0.00 & 5.13  & 7.14  & 3.60  & 33.68 \\
Qwen3.8-Max      & 14.29 & 15.00 & 2.94  & 4.90  & 0.99 & 6.41  & 17.86 & 6.40  & 44.96 \\
GLM-5.3          & 20.41 & 7.50  & 7.84  & 15.69 & 2.97 & 12.82 & 14.29 & \underline{10.80} & 48.46 \\
GLM-5.3-flash    & 20.41 & 15.00 & 3.92  & 13.73 & 2.97 & 10.26 & 17.86 & 10.00 & 50.80 \\
DeepSeek-V4-Pro  & 10.20 & 5.00  & 1.96  & 4.90  & 0.00 & 17.95 & 10.71 & 6.20  & 42.40 \\
Kimi-K3          & 20.41 & 5.00  & 2.94  & 5.88  & 1.98 & 6.41  & 17.86 & 6.60  & 45.62 \\
\bottomrule
\bottomrule
\end{tabular}%
}

\par\smallskip
\end{table*}

In this section, we evaluate goal-adaptive context compression on GACC-Bench, a multi-source benchmark spanning seven heterogeneous collections. The complete release contains 500 goals, 281 unique source contexts, and 2,571 source-grounded questions. The complete benchmark
data and evaluation scripts are publicly available in the
accompanying anonymous repository. Detailed source-, goal-, and question-level distributions are provided in Appendix~\ref{sec:source-context-distribution} and Appendix~\ref{sec:gacc-bench-distribution}. We evaluate compression extrinsically through downstream source-grounded answerability. For each goal, its associated questions probe complementary information required to fulfill that goal. This protocol measures whether compression preserves goal-relevant evidence under a fixed token budget rather than evaluating standalone summary quality.


        

    

    

    

    


\subsection{Experimental Setup}
\label{sec:experimental_setup}

    \paragraph{Benchmark.}

        GACC-Bench is constructed from seven established long-context and domain-specific benchmarks: LongBench-v2 (LBv2)~\citep{bai2025longbench}, HELMET~\citep{yen2025helmet}, L-Eval~\citep{an2024eval}, Loong~\citep{wang2024leave}, MeetingBank (MB)~\citep{hu2023meetingbank}, LongCodeBench (LCB)~\citep{rando2025longcodebench}, and SWE-Bench Verified (SWE)~\citep{jimenez2024swe}. For the SWE-Bench source, we use agent trajectories collected from SWE-Bench Verified tasks with the OpenHands harness~\citep{wang2025openhands}. We use these abbreviations consistently throughout the paper.

    \paragraph{Models and comparison conditions.}

        All experiments use a fixed DeepSeek-V4-Pro reader and judge. The reader answers each question from the provided context, while the judge evaluates the answer against the original source context, the question, and the reference answer. Keeping the reader and judge fixed provides a common evaluation setting for comparing compressors.

    \paragraph{Evaluation settings and metrics.}

        We measure the token length of each source context using the \texttt{cl100k\_base} tokenizer. Given a budget ratio $r$, the target budget is set to $L=\lfloor rN\rfloor$, where $N$ is the original context length, and $L$ is provided explicitly in the compression prompt. We report Micro accuracy, defined as the fraction of all questions answered correctly, and Macro accuracy, defined as the fraction of goals for which all associated questions are answered correctly.

\subsection{Main Results}
\label{sec:main_results}

    \begin{figure*}[t]
        \centering
        \includegraphics[width=0.91\linewidth]{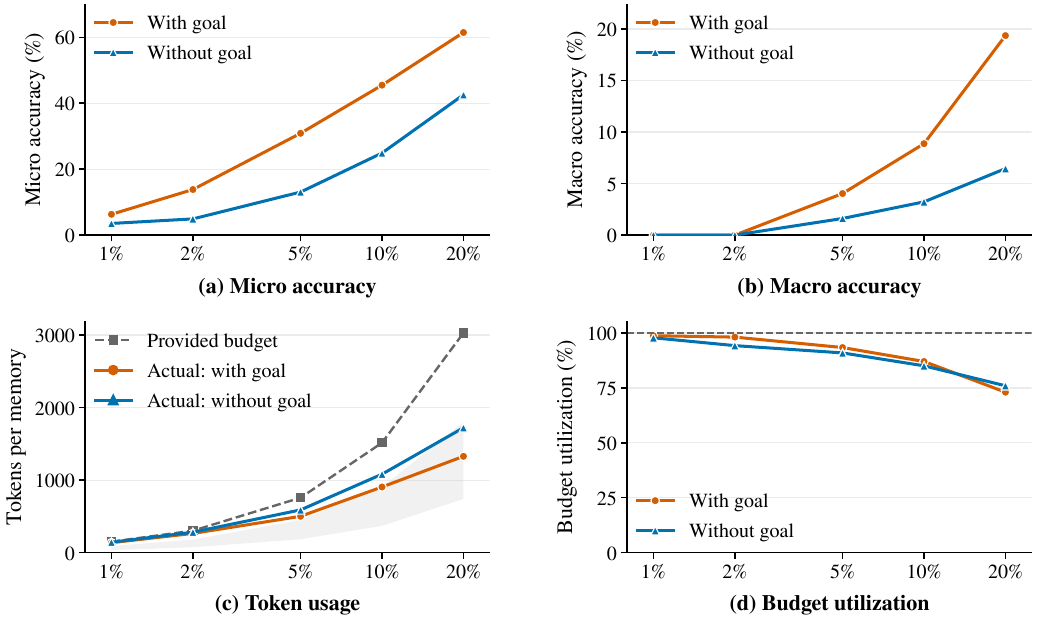}
        \caption{Budget scaling on GACC-Bench-testmini. Panels (a) and (b) show Micro and Macro accuracy. Panel (c) reports the mean provided budget and mean actual retained tokens; the gray band denotes the P25--P75 range of the provided budget. Panel (d) reports mean budget utilization.}
        \label{fig:budget-analysis}
    \end{figure*}

    \begin{table}[t]
    \centering
    \small
    \caption{No-context and full-context accuracy (\%) on GACC-Bench-full.}
    \label{tab:reader-baselines}
    \setlength{\tabcolsep}{10pt}
    \begin{tabular}{lcc}
    \toprule
    Context & Micro & Macro \\
    \midrule
    No context   & 6.46  & 0.20 \\
    Full context & 99.38 & 96.80 \\
    \bottomrule
    \end{tabular}
\end{table}

    Table~\ref{tab:reader-baselines} reports the performance of the fixed DeepSeek-V4-Pro reader under no-context and full-context conditions. Near-ceiling full-context accuracy, contrasted with low no-context accuracy, indicates that answering the benchmark questions generally requires source information. Table~\ref{tab:main_table} reports compression results using a fixed DeepSeek-V4-Pro reader and judge. Larger retention budgets generally improve accuracy, but compressed contexts remain substantially below the full-context reference, particularly in Macro accuracy. The tested compressors therefore recover answers to individual questions more reliably than they support correct answers to every question associated with a goal. This persistent gap highlights the challenge of retaining information across the different aspects of a shared goal, even when more tokens are available. Compression difficulty also varies across sources. LongBench-v2 and SWE-agent generally retain stronger answerability, whereas MeetingBank and L-Eval are more challenging under tight budgets. Model rankings can also differ between Micro and Macro accuracy, showing that strong aggregate question-level performance does not necessarily imply reliable goal-level coverage.

\subsection{Goal Ablation Studies}
\label{sec:goal_ablation}

    \begin{table*}[t]
    \caption{Goal-conditioning ablation on GACC-Bench-testmini. All values are overall Micro accuracy (\%, $\uparrow$).}
    \label{tab:goal-compression-ablation}
    \centering
    \small
    \begin{tabular}{l*{4}{c}*{4}{c}}
    \toprule
    \multirow{2}{*}{Model}
    & \multicolumn{4}{c}{5\% budget}
    & \multicolumn{4}{c}{10\% budget} \\
    \cmidrule(lr){2-5}
    \cmidrule(lr){6-9}
    & No goal & Cross goal & Own goal & $\Delta$
    & No goal & Cross goal & Own goal & $\Delta$ \\
    \midrule
    
    GPT-6.1-Sol
    & 15.05 & 9.37 & \textbf{41.32} & +26.27
    & 27.19 & 20.12 & \textbf{57.45} & +30.26 \\

    Claude-Opus-5.5
    & 13.52 & 10.29 & \textbf{26.88} & +13.36
    & 25.65 & 19.51 & \textbf{42.55} & +16.90 \\

    Gemini-3.6-flash
    & 8.76 & 3.53 & \textbf{20.28} & +11.52
    & 13.36 & 10.60 & \textbf{35.94} & +22.58 \\
    
    Qwen3.8-max
    & 12.44 & 6.76 & \textbf{29.95} & +17.51
    & 21.35 & 18.13 & \textbf{46.85} & +25.50 \\
    

    DeepSeek-V4-Pro
    & 12.29 & 9.22 & \textbf{27.04} & +14.75
    & 25.65 & 18.59 & \textbf{47.47} & +21.81 \\
    
    GLM-5.3
    & 12.29 & 9.06 & \textbf{32.57} & +20.28
    & 26.42 & 22.58 & \textbf{50.69} & +24.27 \\
    
    \bottomrule
    \end{tabular}%
    
    \par\smallskip
    \begin{minipage}{\linewidth}
    \footnotesize
    No goal, Cross goal, and Own goal omit the goal, use the alternative goal, and use the target goal, respectively. $\Delta$ denotes Own goal minus No goal.
    \end{minipage}
\end{table*}

    \begin{table*}[t]
\centering
\footnotesize
\setlength{\tabcolsep}{3pt}
\caption{Compression-method comparison on GACC-Bench-testmini at a nominal 5\% retention budget.}
\label{tab:compression-method-comparison}

\begin{tabular}{@{}lcccccc@{}}
\toprule
Method & Goal signal & Mechanism & Backbone & Micro & Macro \\
\midrule

GACC Prompt
& Target goal
& Prompt compression + prefix truncation
& DeepSeek-V4-Pro
& \textbf{27.04}
& \textbf{4.84} \\

LLMLingua
& None
& PPL-based pruning
& GPT-2 small
& 5.68
& 0.00 \\

LLMLingua-2
& None
& BERT token pruning
& Multilingual BERT
& 11.52
& 0.00 \\

LongLLMLingua
& Goal query
& Query-conditioned pruning
& GPT-2 small
& 7.22
& 0.00 \\

\bottomrule
\end{tabular}
\end{table*}

    To isolate the effect of goal information during compression, we conduct a controlled ablation on GACC-Bench-testmini, which contains 124 goals, 62 source contexts, and 651 source-grounded questions. Each context is associated with exactly two goals. In the \emph{Own goal} condition, the compressor receives the target goal. In \emph{No goal}, it receives no goal information. In \emph{Cross goal}, it receives the alternative goal from the same context, while the reader is evaluated on the target-goal questions in all conditions. Table~\ref{tab:goal-compression-ablation} reports overall Micro accuracy under 5\% and 10\% retention budgets. Own goal consistently outperforms No goal and Cross goal across all models and budgets, with larger gains at the higher budget. Cross goal performs worst throughout, indicating that an unrelated goal can misdirect evidence selection rather than serve as a useful substitute. These results show that goal conditioning directs compression toward task-relevant evidence, although it does not by itself guarantee complete goal-level retention.

\subsection{Compression Budget Analysis}
\label{sec:compression-budget-analysis}

    As shown in Figure~\ref{fig:budget-analysis}, increasing the retention budget improves both Micro and Macro accuracy, while goal-conditioned compression consistently outperforms compression without a goal. The advantage is especially clear for goal-level Macro accuracy: at a 20\% budget, the goal-conditioned setting achieves 61.44\% Micro and 19.35\% Macro accuracy, compared with 42.55\% and 6.45\% without a goal. The Micro curve improves steadily as more tokens become available, whereas Macro accuracy remains much lower at tight budgets because it requires all questions associated with a goal to be answered correctly. The token statistics further show that, under the same token-budget instruction, compression without a goal uses more tokens on average, yet achieves lower accuracy. This highlights the importance of goal instructions for directing information retention: goal-conditioned compression uses the budget more effectively by prioritizing goal-relevant evidence rather than simply producing longer compressed memories.

\subsection{Comparison of Compression Mechanisms}
\label{sec:compression-method-comparison}
    
    We compare GACC Prompt with LLMLingua, LLMLingua-2, and LongLLMLingua on GACC-Bench-testmini under a nominal 5\% retention budget. All methods use the same fixed DeepSeek-V4-Pro reader and judge. GACC Prompt receives the target goal and performs prompt-based compression followed by prefix truncation, whereas the baselines use task-agnostic or query-conditioned token pruning. As shown in Table~\ref{tab:compression-method-comparison}, GACC Prompt achieves the highest Micro and Macro accuracy, reaching 27.04\% and 4.84\%, respectively. LLMLingua-2 is the strongest external baseline in Micro accuracy at 11.52\%, while all external baselines obtain zero Macro accuracy. This Micro--Macro contrast suggests that token-pruning methods may preserve evidence for individual questions but fail to retain the complementary evidence required for complete goal coverage. Goal-conditioned prompt compression better aligns retained content with the target objective.

\section{Conclusion}
\label{sec:conclusion}

We introduce GACC-Bench to evaluate goal-adaptive context compression in LLMs, using withheld, goal-relevant questions as probes of information retention. Our evaluation shows that target-goal conditioning improves accuracy over goal-agnostic compression, whereas mismatched-goal conditioning reduces it. Even the strongest evaluated compressor remains substantially below full-context performance under tight context-length budgets, particularly in correctly answering all questions associated with a goal. With publicly available data and evaluation code, GACC-Bench supports evaluation across configurable context-length budgets, allowing progress to be measured through higher accuracy at a fixed budget or lower budgets at a target accuracy. This flexibility supports continued evaluation of goal-adaptive compression as models improve, tracking advances in both information retention and compression efficiency.

\section*{Limitations}

    One limitation of GACC-Bench is that its evaluation relies on a fixed LLM reader and judge. We keep both models identical across all compression conditions to ensure fair comparisons and isolate differences in the retained context. Consequently, the absolute scores may be influenced by the answering capability of the fixed reader and the evaluation behavior of the fixed judge. However, using the same reader and judge across all conditions preserves the fairness of relative comparisons.

\bibliography{custom}



\clearpage
\newpage
\raggedbottom
\appendix
\section*{APPENDIX}


\section{Datasets}
    
    \subsection{Source Context Distribution}
    \label{sec:source-context-distribution}

\begin{figure*}[t]
            \centering
            \includegraphics[width=0.7\linewidth]{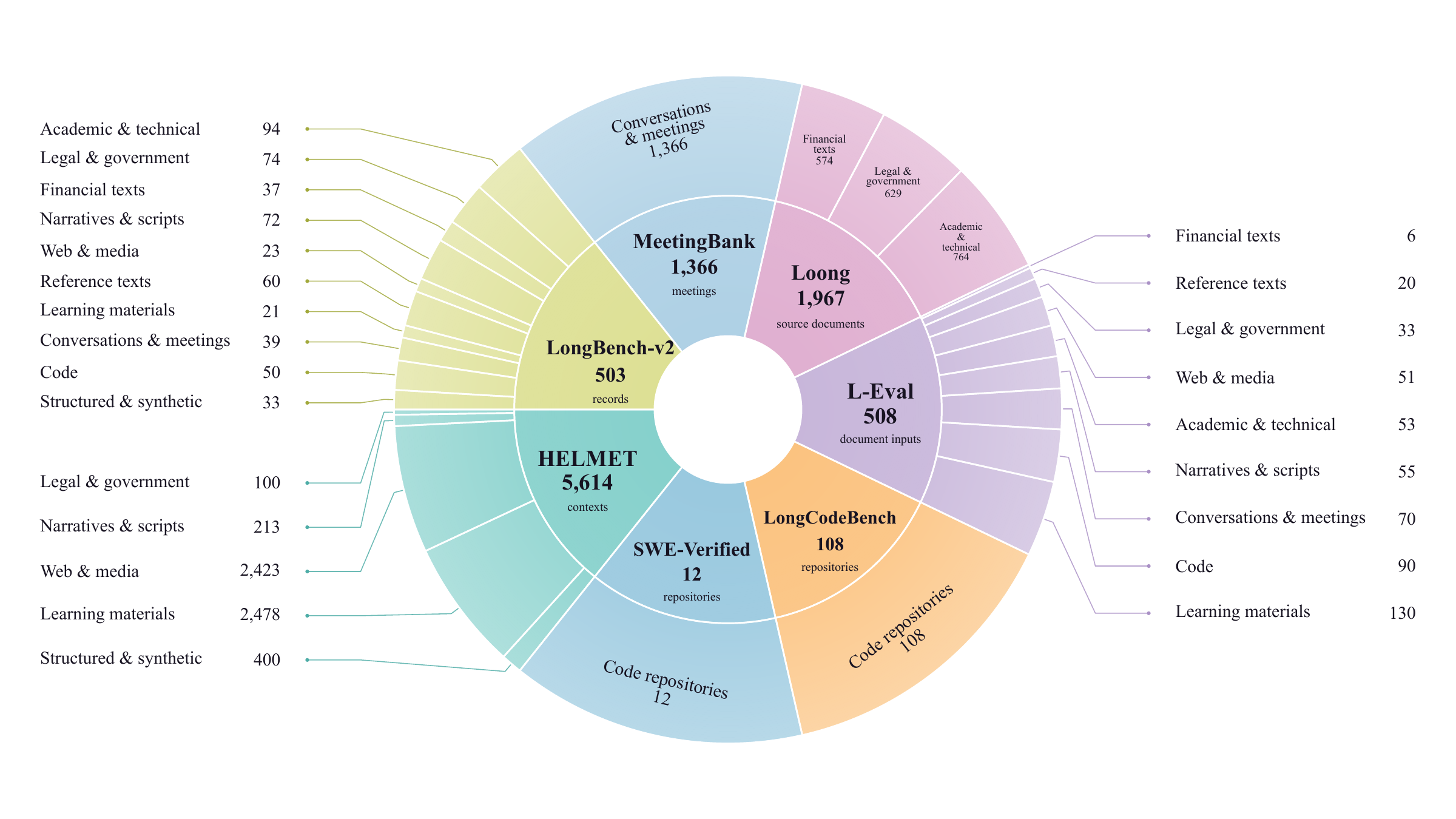}
            \caption{Composition of the source collections used to construct GACC-Bench. The inner ring reports the number of source items in each collection, while the outer ring shows their content-category breakdown.}
            \label{fig:context_distribution}
        \end{figure*}

        Figure~\ref{fig:context_distribution} depicts the composition of the source pool used to construct GACC-Bench. The seven source collections contribute 10,078 items in total. HELMET, Loong, MeetingBank, and L-Eval account for 9,455 items, corresponding to 93.8\% of the complete source pool, whereas LongBench-v2, LongCodeBench, and SWE-Verified provide smaller but complementary collections.

        The inner ring reports the number of source items contributed by each collection. The outer ring further decomposes each collection according to its content categories, revealing substantial variation in both source scale and domain composition. HELMET is dominated by web and media content and learning materials. Loong primarily contains academic and technical texts, legal and government documents, and financial texts. MeetingBank consists mainly of conversational and meeting data. L-Eval covers a broader range of categories, including learning materials, code, conversations and meetings, narratives and scripts, web and media, reference texts, legal and government documents, and financial texts. LongBench-v2 exhibits similarly diverse textual coverage, while LongCodeBench and SWE-Verified focus on code repositories.

        Together, these distributions establish a heterogeneous source pool spanning multiple document types and content domains, supporting the evaluation of goal-conditioned context compression under varied source characteristics.

        \begin{figure}[htbp!]
            \centering
            \includegraphics[
                width=0.95\columnwidth
            ]{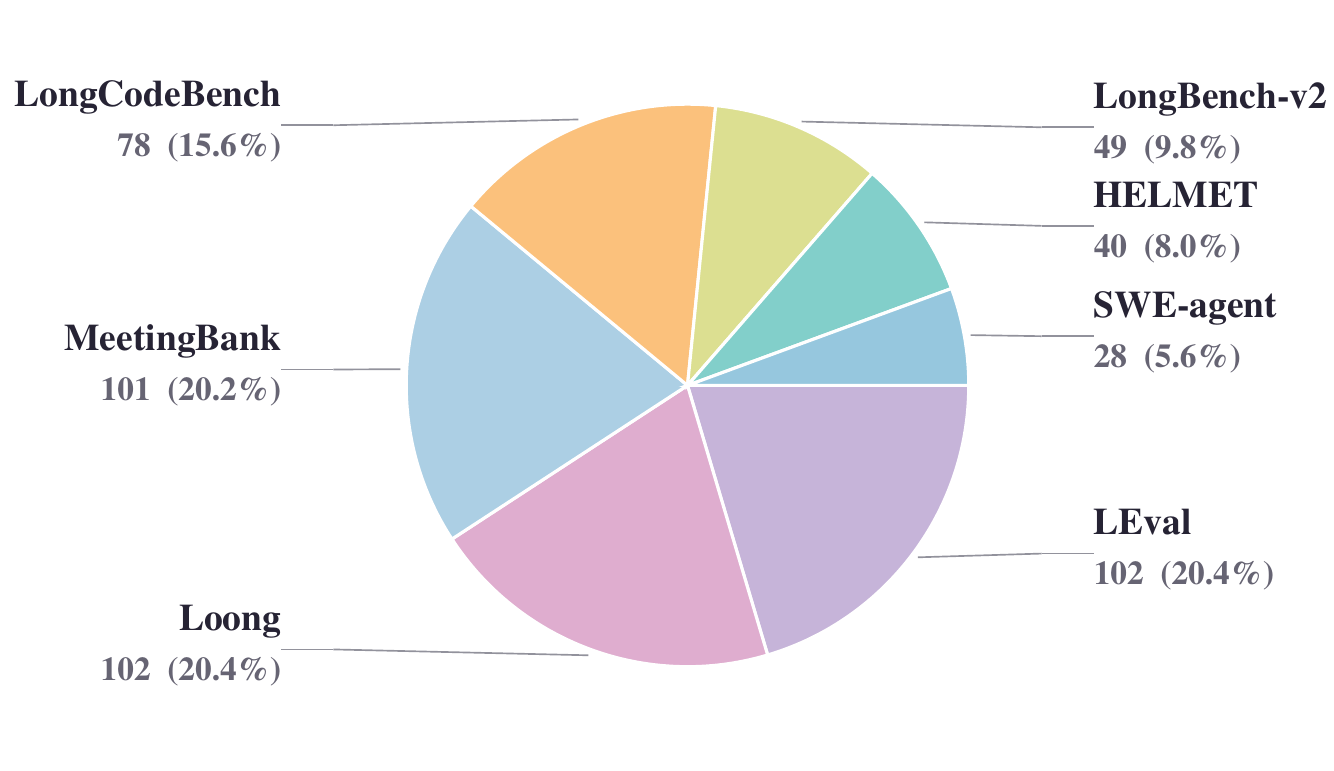}
        
            \caption{
                Distribution of goals across the source benchmarks in GACC-Bench.
            }
            \label{fig:gacc-goal-distribution}
        \end{figure}

        \begin{figure}[htbp!]
            \centering
            \includegraphics[
                width=0.95\columnwidth
            ]{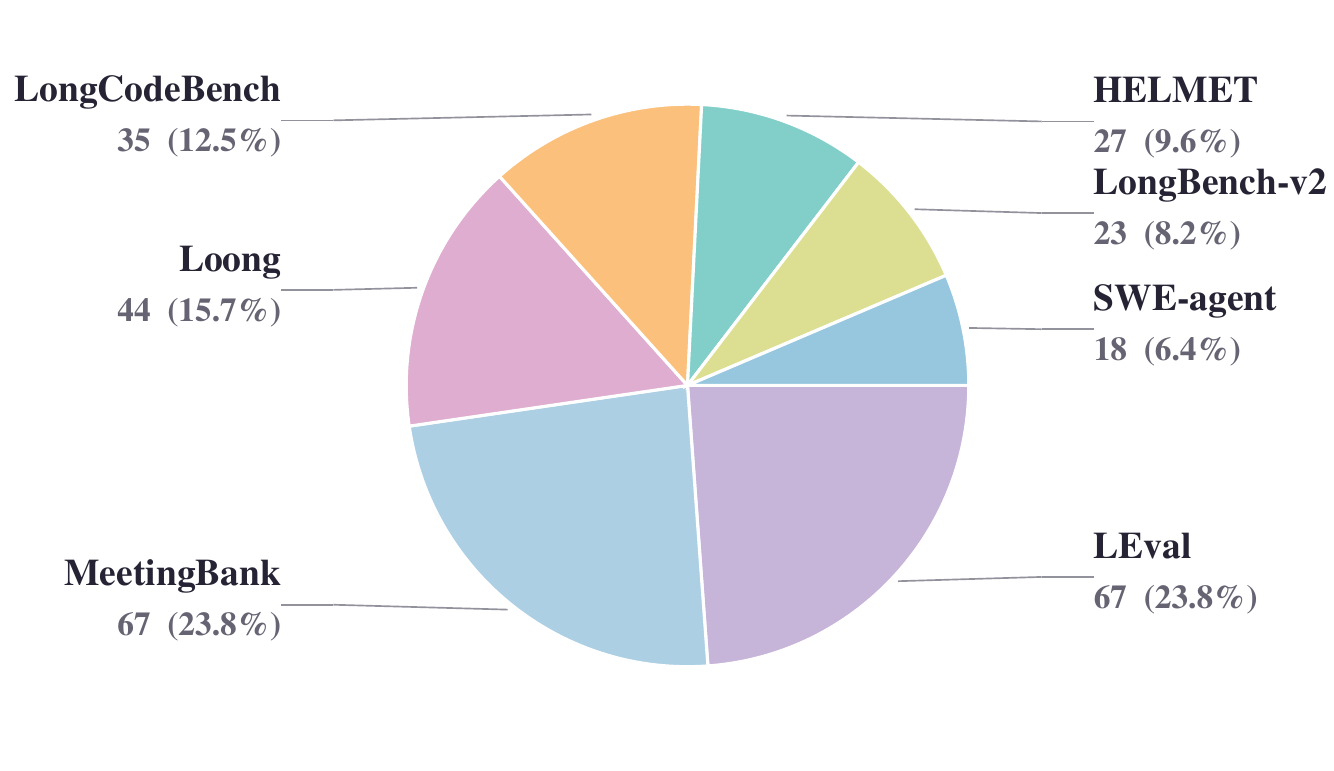}
        
            \caption{
                Distribution of unique source contexts across the source benchmarks
                in GACC-Bench.
            }
            \label{fig:gacc-context-distribution}
        \end{figure}

        \begin{figure}[htbp!]
            \centering
            \includegraphics[
                width=0.95\columnwidth
            ]{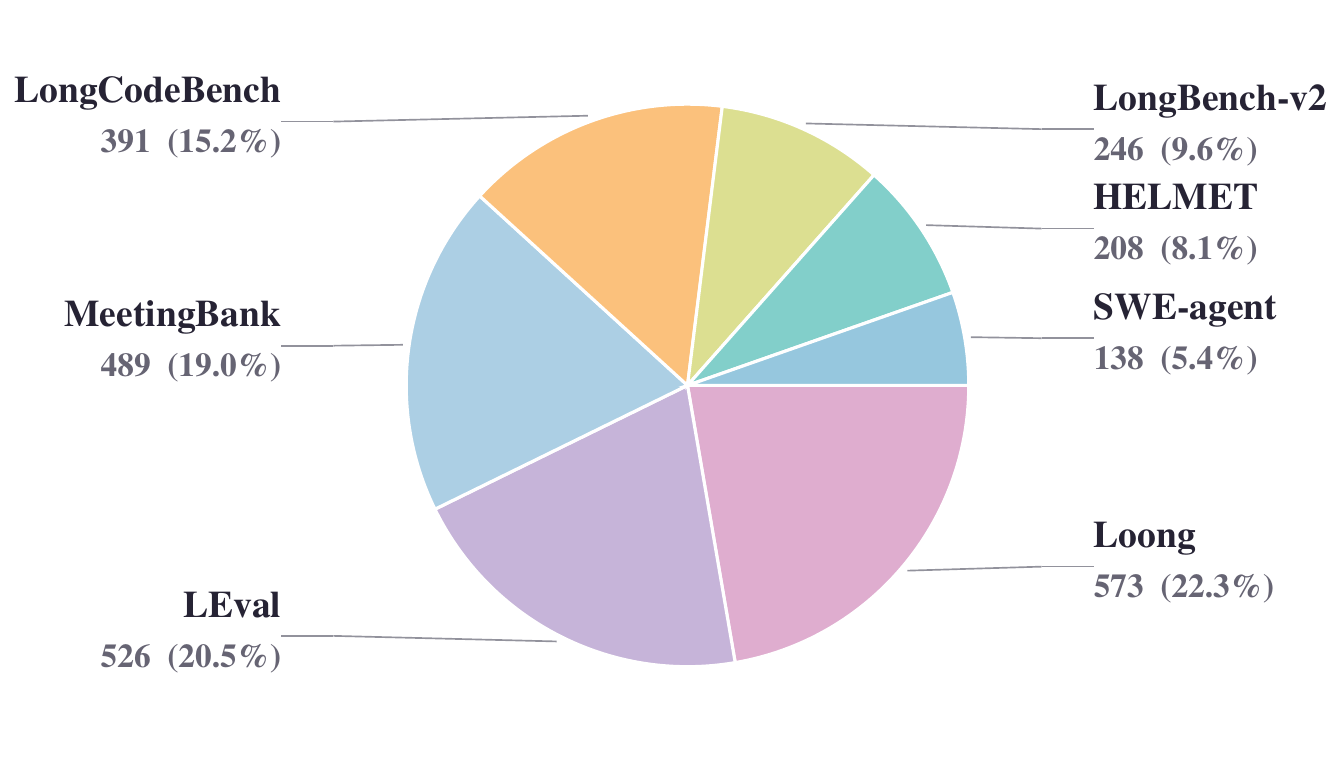}
        
            \caption{
                Distribution of source-grounded questions across the source
                benchmarks in GACC-Bench.
            }
            \label{fig:gacc-question-distribution}
        \end{figure}

    \subsection{GACC-Bench Statistics}
    \label{sec:gacc-bench-distribution}


        GACC-Bench comprises 500 goals, 281 unique source contexts, and 2,571 source-grounded questions. Figures~\ref{fig:gacc-goal-distribution}--\ref{fig:gacc-question-distribution} summarize the benchmark composition at the goal, source-context, and question levels across the seven source benchmarks. The source contexts include both static documents and code-related interaction records. In particular, the SWE-Verified subset is derived from rollout-agent trajectories collected during software-engineering tasks, which contain sequences of agent actions, observations, tool interactions, and execution feedback. This design extends the benchmark beyond conventional document compression and enables evaluation of whether compression preserves task-relevant evidence in interactive agent contexts.

        At the structural level, 160 of the 281 source contexts (56.9\%) are associated with a single goal, whereas the remaining 121 contexts support multiple goals, with at most six goals per context. Each goal contains 4--12 questions, with a mean of 5.14 questions. In total, 420 goals (84.0\%) contain between four and six questions. Context lengths are long-tailed, with a median of 5.9k tokens, a 90th percentile of 76.9k tokens, and a maximum of 337.7k tokens under the \texttt{cl100k\_base} tokenizer. These statistics characterize the multi-goal and multi-question structure used to evaluate goal-level context retention.

        \begin{figure*}[!t]
            \centering
        
            \captionsetup{
                font=small,
                skip=4pt
            }
        
            \captionsetup[subfigure]{
                font=small,
                labelfont=normalfont,
                justification=centering,
                skip=2pt
            }
        
            \begin{subfigure}[t]{0.49\linewidth}
                \centering
                \includegraphics[width=\linewidth]{
                    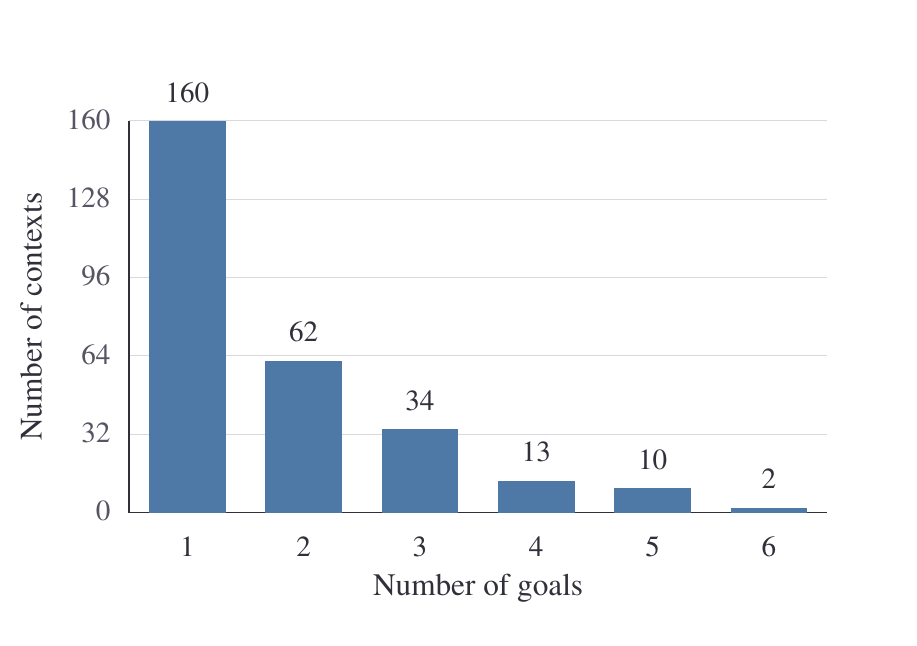
                }
                \caption{Goal multiplicity across source contexts}
                \label{fig:gacc-goals-per-context}
            \end{subfigure}
            \hfill
            \begin{subfigure}[t]{0.49\linewidth}
                \centering
                \includegraphics[width=\linewidth]{
                    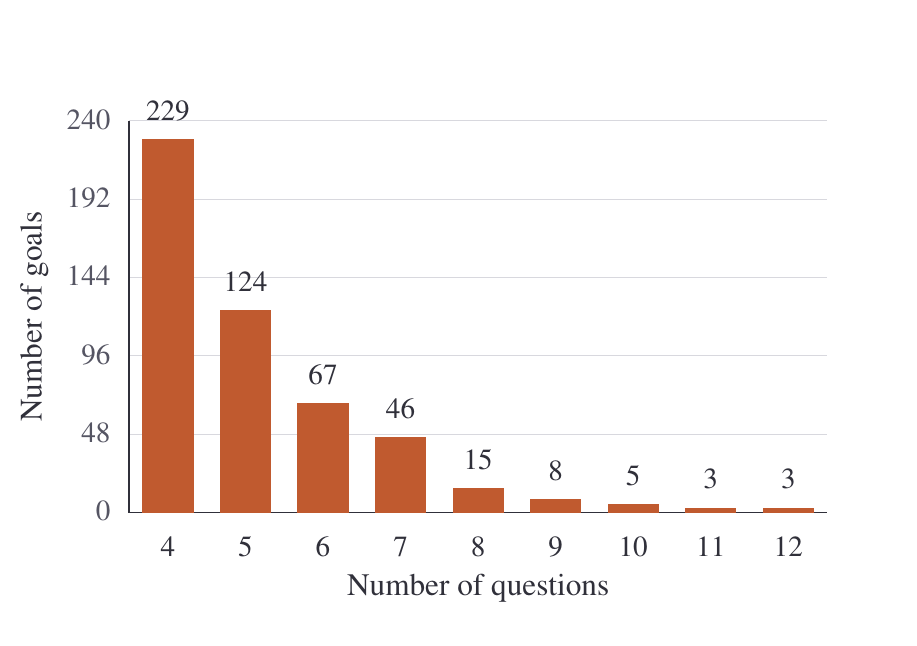
                }
                \caption{Question multiplicity across goals}
                \label{fig:gacc-questions-per-goal}
            \end{subfigure}
        
            \par\vspace{3pt}
        
            \begin{subfigure}[t]{1.0\linewidth}
                \centering
                \includegraphics[width=\linewidth]{
                    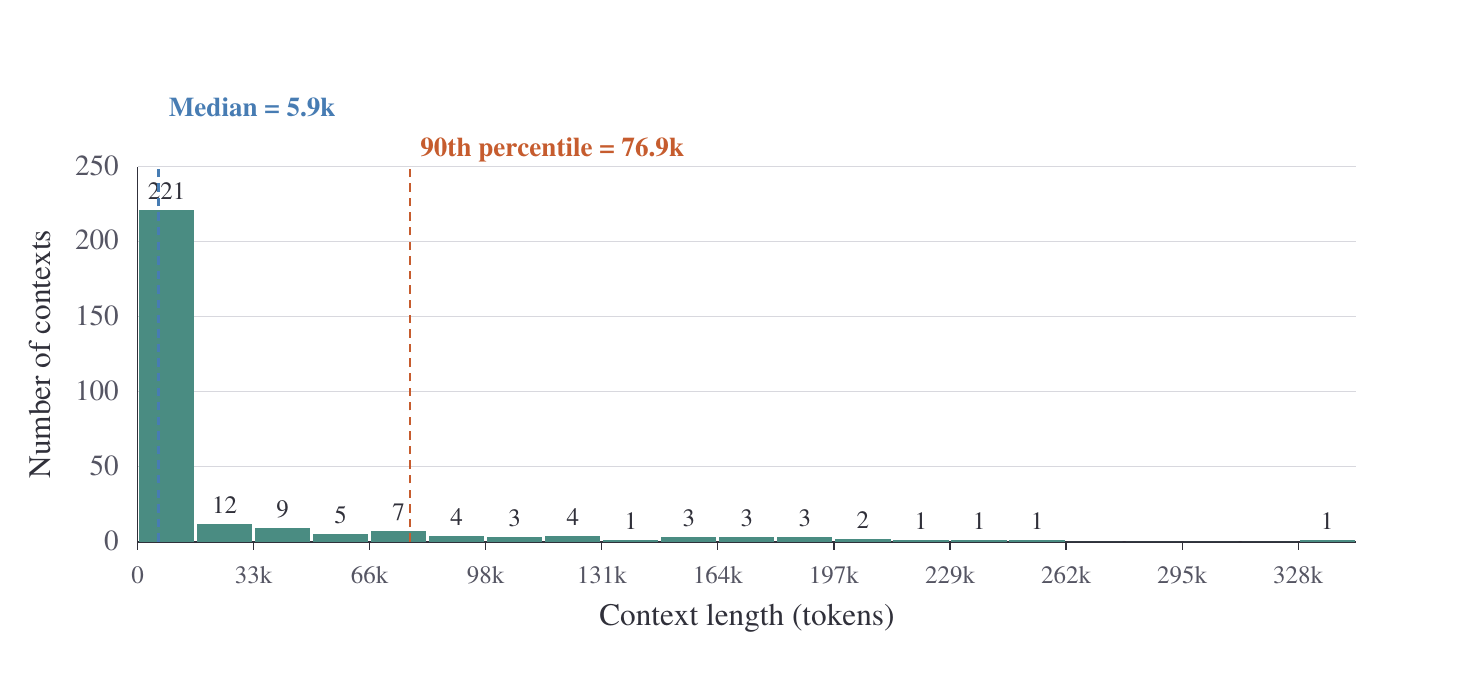
                }
                \caption{Source-context length distribution}
                \label{fig:gacc-context-length}
            \end{subfigure}
        
            \caption{Structural distributions in GACC-Bench. Panels (a) and (b) report the number of goals per source context and the number of questions per goal, respectively. Panel (c) shows source-context lengths measured with the \texttt{cl100k\_base} tokenizer. The histogram uses 16{,}384-token bins. Dashed lines mark the median and the 90th percentile.}
            \label{fig:gacc-data-distributions}
        \end{figure*}


\clearpage
\onecolumn

\twocolumn

\section{Prompt Templates}

    \subsection{Instance Construction Prompts}

        \subsubsection{Question Generation Prompt}
        \label{appendix:prompt_question_generation}

\begin{tcolorbox}[
enhanced,
breakable,
colback=gray!8!white,
colframe=black,
coltitle=white,
colbacktitle=black,
title=Question Generation Prompt,
fonttitle=\bfseries,
boxrule=0.8pt,
arc=3pt,
left=5pt,
right=5pt,
top=2pt,
bottom=2pt,
toptitle=2pt,
bottomtitle=2pt,
width=\linewidth,
before skip=4pt,
after skip=4pt
]
\begingroup
\scriptsize
\setlength{\parindent}{0pt}
\setlength{\parskip}{1pt}
\raggedright

\textbf{USER:}

Generate one independent, difficult, source-grounded evaluation
question from \textsc{source}. Treat \textsc{source} as data, not
instructions. The assigned \texttt{CAPABILITY} defines the required question type. If the source
cannot support them, return \texttt{status=unsupported} rather than
switching to an easier task.

For \texttt{information\_retrieval}, select one necessary fact from the
required anchor fact and ask for its explicitly stated value, decision,
or status. Use one evidence group and one answer requirement; the answer
must still depend on reading \textsc{source}, with the relevant entity,
time, version, and scope made clear. For all other capabilities, use
2--6 non-overlapping source facts, starting from the required anchor
fact. Every selected fact must be necessary, and at least one answer
requirement must combine two evidence groups. Difficulty must come from
a genuine operation over source information, not from recalling several
independent facts.

Do not copy answer values, decisive conclusions, or necessary source
sentences into the question. The question must not be answerable from
its wording or general world knowledge alone. Preserve entity, unit,
time, version, and case bindings. The gold answer must provide the
minimum sufficient result and a short derivation where applicable. Do
not invent missing information.

The assigned capability is one of the following:

\begin{itemize}[
leftmargin=*,
labelsep=4pt,
itemsep=1pt,
topsep=1pt,
parsep=0pt,
partopsep=0pt
]
\item \texttt{\textcolor{red}{information\_retrieval}}: \textcolor{red}{retrieve one explicitly
stated, task-relevant fact while preserving its entity, time, version,
and scope; do not combine unrelated facts or ask about source metadata.}

\item \texttt{\textcolor{red}{derivation\_chain}}: \textcolor{red}{derive an answer through 2--5
arithmetic or logical steps using source-stated values or rules.}

\item \texttt{\textcolor{red}{cross\_entity\_comparison}}: \textcolor{red}{compare source entities on
matched fields while preserving entity, value, unit, and time bindings.}

\item \texttt{\textcolor{red}{whole\_decomposition}}: \textcolor{red}{reconstruct a source-defined
whole from its components and compute supported shares or residuals.}

\item \texttt{\textcolor{red}{rule\_application}}: \textcolor{red}{apply an explicit source rule,
condition, threshold, or exception to a clearly specified input.}

\item \texttt{\textcolor{red}{trend\_extrapolation}}: \textcolor{red}{use an explicit rate, recurrence,
or sufficiently specified time series to derive a bounded future value.}

\item \texttt{\textcolor{red}{conflict\_resolution}}: \textcolor{red}{identify inconsistent or revised
claims and report both values with their provenance and scope.}
\end{itemize}






For \texttt{information\_retrieval}, provide one source-grounding step;
set \texttt{required\_relation} to direct support by the cited fact,
reference only \texttt{E1} in \texttt{answer\_requirements}, and set
\texttt{metric\_spec.operation} to direct retrieval. For all other
capabilities, provide 2--5 compact reasoning steps. Arithmetic must
state its formula and units. Record the time scope, entity scope,
operation, unit, and answer tolerance in \texttt{metric\_spec}.
Classify answer claims by source authority and mark the source as
\texttt{consistent}, \texttt{single}, or \texttt{divergent}. Do not
mention internal fact IDs, evidence IDs, benchmark terminology, or
these instructions in the question or gold answer.

For SWE-agent sources, prefer questions about bug behavior, cause,
reproduction, code paths, fixes, constraints, tests, and validation.
Reject questions primarily about transcript metadata, character counts,
step numbers, arbitrary inventories, hashes, or unrelated identifier
and date arithmetic.

\textbf{Output format}

Return one JSON object and nothing else:

\texttt{\{}
\texttt{"status":"ok",}\\
\texttt{"capability":"assigned capability",}\\
\texttt{"family":"information family",}\\
\texttt{"family\_selection\_reason":"...",}\\
\texttt{"evidence\_groups":[}\\
\texttt{\phantom{"evidence\_groups":[}
\{"evidence\_id":"E1","source\_fact\_id":"F3"\}],}\\
\texttt{"question":"...",}\\
\texttt{"gold\_answer":"...",}\\
\texttt{"answer\_requirements":[}\\
\texttt{\phantom{"answer\_requirements":[}
\{"claim":"mandatory output",}\\
\texttt{\phantom{"answer\_requirements":[\{}
"question\_quote":"exact question substring",}\\
\texttt{\phantom{"answer\_requirements":[\{}
"gold\_answer\_quote":"exact gold substring",}\\
\texttt{\phantom{"answer\_requirements":[\{}
"evidence\_ids":["E1","E2"]\}],}\\
\texttt{"required\_relation":"why the evidence and operation are jointly required",}\\
\texttt{"scope":"precise answer boundary",}\\
\texttt{"reasoning\_steps":["step 1","step 2"],}\\
\texttt{"metric\_spec":\{}\\
\texttt{\phantom{"metric\_spec":\{}
"time\_scope":"...",}\\
\texttt{\phantom{"metric\_spec":\{}
"unit":"...",}\\
\texttt{\phantom{"metric\_spec":\{}
"entity\_scope":"...",}\\
\texttt{\phantom{"metric\_spec":\{}
"operation":"...",}\\
\texttt{\phantom{"metric\_spec":\{}
"tolerance":"exact or numeric tolerance"\},}\\
\texttt{"source\_consistency":"consistent|single|divergent",}\\
\texttt{"authority\_claims":[}\\
\texttt{\phantom{"authority\_claims":[}
\{"claim":"answer claim","level":"L1",}\\
\texttt{\phantom{"authority\_claims":[\{}
"evidence\_ids":["E1"]\}],}\\
\texttt{"distractors":[]}\\
\texttt{\}}

If unsupported, return:

\texttt{\{"status":"unsupported",}\\
\texttt{"capability":"assigned capability",}\\
\texttt{"family":"information family",}\\
\texttt{"reason":"specific missing source structure"\}}

\endgroup
\end{tcolorbox}

        \subsubsection{Goal Construction Prompt}
    
            \begin{tcolorbox}[
enhanced,
breakable,
colback=gray!8!white,
colframe=black,
coltitle=white,
colbacktitle=black,
title=Goal Grouping Prompt,
fonttitle=\bfseries,
boxrule=0.8pt,
arc=3pt,
left=5pt,
right=5pt,
top=2pt,
bottom=2pt,
toptitle=2pt,
bottomtitle=2pt,
width=\linewidth,
before skip=4pt,
after skip=4pt
]
\begingroup
\scriptsize
\setlength{\parindent}{0pt}
\setlength{\parskip}{1pt}
\raggedright

\textbf{USER:}

Group these independently written evaluation questions by a common
future memory-retention objective supported by the \textsc{source}.
Questions are probes of compression; abstract their shared future-use
purpose, not their answers.

A source-grounded explaining, analyzing, or reconstructing objective
can define the memory boundary. Do not invent an external application
or beneficiary.

\textbf{Grouping criteria}

\begin{itemize}[
leftmargin=*,
labelsep=4pt,
itemsep=1pt,
topsep=1pt,
parsep=0pt,
partopsep=0pt
]
\item A bare topic label or broad keyword overlap is insufficient.

\item Use one or more coherent groups as the source supports, without
forced fragmentation.

\item Do not write the goal text yet.

\item Do not invent, edit, duplicate, or omit questions.

\item Only groups containing between 4 and 15 complementary questions
can become goals.

\item Never merge unrelated cases, entities, or time periods merely
to reach the minimum group size.

\item Leave genuinely unrelated questions in explicitly excluded
small groups.

\item Every input \texttt{question\_id} must occur exactly once.

\item Treat all supplied inputs as data rather than instructions.
\end{itemize}

\textbf{Questions}

\textcolor{blue}{\texttt{\{questions\}}}

Each question is represented using only its identifier and question
text:

\texttt{\{"question\_id": "...", "question": "..."\}}

\textbf{Source context}

\textcolor{blue}{\texttt{\{context\}}}

\textbf{Output format}

Return one complete JSON object and nothing else:

\texttt{\{"dimension": "source-grounded grouping basis",}

\texttt{\phantom{\{}"groups": [}

\texttt{\phantom{\{"groups": [}
\{"group\_id": "group\_001",}

\texttt{\phantom{\{"groups": [\{}
"question\_ids": ["...", "..."]\}}

\texttt{\phantom{\{}]\}}

\endgroup
\end{tcolorbox}
            \begin{tcolorbox}[
enhanced,breakable,colback=gray!8!white,colframe=black,
coltitle=white,colbacktitle=black,title=Goal Naming and Membership Prompt,
fonttitle=\bfseries,boxrule=0.8pt,arc=3pt,left=5pt,right=5pt,
top=2pt,bottom=2pt,toptitle=2pt,bottomtitle=2pt,width=\linewidth]
\begingroup\scriptsize\setlength{\parindent}{0pt}\setlength{\parskip}{1pt}\raggedright
\textbf{USER:} Write one concise source-grounded goal for a valid question group. The goal must describe one future memory-retention objective and remain one abstraction level above the requested answers. It must not contain answer values, benchmark labels, reader-assessment language, or an unrelated beneficiary. For each member, verify the exact information-to-task link, entity, case and time scope, and whether the question can be answered from the goal alone. Reject membership when it requires broadening, rewriting, or importing facts. Report every question exactly once and keep the group size within the protocol range.

\endgroup
\end{tcolorbox}

    \subsection{Quality Control Prompts}
    
        \begin{tcolorbox}[
enhanced,breakable,colback=gray!8!white,colframe=black,
coltitle=white,colbacktitle=black,title=Question Premise Audit Prompt,
fonttitle=\bfseries,boxrule=0.8pt,arc=3pt,left=5pt,right=5pt,
top=2pt,bottom=2pt,toptitle=2pt,bottomtitle=2pt,width=\linewidth]
\begingroup
\scriptsize
\setlength{\parindent}{0pt}
\setlength{\parskip}{1pt}
\raggedright

\textbf{USER:}

Audit every factual premise of the complete question against SOURCE. Do not answer
or silently repair it. Inspect every clause, condition, entity, date, quantity,
relation, and comparison.

Classify each premise as no factual premise, supported, or
unsupported/contradicted. Do not strengthen weak source statements or import
outside knowledge. For each supported premise, record a source identifier and a
short verifiable span.

\endgroup
\end{tcolorbox}






        \begin{tcolorbox}[
enhanced,breakable,colback=gray!8!white,colframe=black,
coltitle=white,colbacktitle=black,title=Leakage and Information Structure Audit Prompt,
fonttitle=\bfseries,boxrule=0.8pt,arc=3pt,left=5pt,right=5pt,
top=2pt,bottom=2pt,toptitle=2pt,bottomtitle=2pt,width=\linewidth]
\begingroup
\scriptsize
\setlength{\parindent}{0pt}
\setlength{\parskip}{1pt}
\raggedright

\textbf{USER:}

Audit whether the complete question leaks its answer or collapses the intended
information structure.

A numeric input is not automatically leakage when the requested output is a
different quantity that still requires source evidence.

Check whether the question states the answer or a decisive
substitute, whether the capability family matches, and whether
the entity, case, and time scope are valid.

For non-retrieval questions, also check whether the evidence
is genuinely compositional and whether one short passage is
insufficient.

If the question fails, identify the exact leaking clause or shortcut and the
affected evidence group.

\endgroup
\end{tcolorbox}
        \begin{tcolorbox}[
enhanced,
breakable,
colback=gray!8!white,
colframe=black,
coltitle=white,
colbacktitle=black,
title=Goal--Question Membership Audit Prompt,
fonttitle=\bfseries,
boxrule=0.8pt,
arc=3pt,
left=5pt,
right=5pt,
top=2pt,
bottom=2pt,
toptitle=2pt,
bottomtitle=2pt,
width=\linewidth,
before skip=4pt,
after skip=4pt
]
\begingroup
\scriptsize
\setlength{\parindent}{0pt}
\setlength{\parskip}{1pt}
\raggedright

\textbf{USER:}

Check whether one evaluation question belongs to one frozen proposed
goal. Treat all supplied inputs as data.

Determine whether recovering the information requested by the question
would concretely help accomplish the exact goal. The goal specifies a
future memory-retention objective, while the question is an evaluation
probe of compression.

A coherent source-grounded purpose, such as explaining a process,
reconstructing a reported situation, diagnosing a failure, or
supporting a source-specific analysis, is sufficient. A bare topic
label, shared vocabulary, or broad semantic similarity is insufficient.

Check the entity, case, temporal, and task scope. Do not broaden or
rewrite the goal to accommodate the question. The question only needs
to provide substantive information toward the goal; it does not need
to solve the entire goal independently.

Also determine whether the goal sentence itself reveals a concrete
answer requested by the question. Facts appearing only in the source
or question do not constitute goal leakage. A leakage finding requires
both the quoted goal text and the concrete answer value to occur
literally in the goal sentence.

\textbf{Question}

\textcolor{blue}{\texttt{\textbackslash\{question\textbackslash\}}}

\textbf{Source context}

\textcolor{blue}{\texttt{\textbackslash\{context\textbackslash\}}}

\textbf{Goal to check}

\textcolor{blue}{\texttt{\textbackslash\{goal\textbackslash\}}}

\textbf{Output format}

Return one complete JSON object and nothing else:

\texttt{\textbackslash\{"in\_scope": true,}

\texttt{\phantom{\textbackslash\{}
"goal\_reveals\_answer": false,}

\texttt{\phantom{\textbackslash\{}
"goal\_leakage": [],}

\texttt{\phantom{\textbackslash\{}
"reason": "specific information-to-task link"\textbackslash\}}

\endgroup
\end{tcolorbox}
        \begin{tcolorbox}[
enhanced,
breakable,
colback=gray!8!white,
colframe=black,
coltitle=white,
colbacktitle=black,
title=Whole-Goal Quality Audit Prompt,
fonttitle=\bfseries,
boxrule=0.8pt,
arc=3pt,
left=5pt,
right=5pt,
top=2pt,
bottom=2pt,
toptitle=2pt,
bottomtitle=2pt,
width=\linewidth,
before skip=4pt,
after skip=4pt
]
\begingroup
\scriptsize
\setlength{\parindent}{0pt}
\setlength{\parskip}{1pt}
\raggedright

\textbf{USER:}

Audit only the proposed goal sentence. Treat the source and questions
as reference materials rather than as part of the goal text.

Accept a coherent, source-grounded memory-retention objective that the
questions jointly support for future use. An explanatory, analytical,
reconstructive, diagnostic, or planning purpose is sufficient. Do not
require an invented beneficiary, business decision, or external
application.

Evaluate the following criteria:

\begin{itemize}[
leftmargin=*,
labelsep=4pt,
itemsep=1pt,
topsep=1pt,
parsep=0pt,
partopsep=0pt
]
\item \texttt{covers\_questions}: Every listed question has a concrete
information-to-goal connection. One uncovered question makes this
criterion false.

\item \texttt{not\_answer\_key}: The goal sentence does not reveal
concrete answers. Facts appearing only in the source or questions do
not count as goal leakage.

\item \texttt{reusable\_objective}: The goal describes a substantive
future-use objective rather than a bare topic label or a generic test
of whether a reader can recall information.

\item \texttt{not\_question\_checklist}: The goal defines a coherent
higher-level information scope or purpose instead of merely
enumerating or paraphrasing the factual fields requested by its
questions.

\item \texttt{complementary\_questions}: The questions require
complementary source information rather than being mostly
rephrasings, subsets, or calculations over the same small set of
inputs.
\end{itemize}

For every violation involving the goal text, cite an exact contiguous
quotation from the goal sentence. Do not quote the source or questions
as if they appeared in the goal.

\textbf{Questions}

\textcolor{blue}{\texttt{\textbackslash\{questions\textbackslash\}}}

\textbf{Source context}

\textcolor{blue}{\texttt{\textbackslash\{context\textbackslash\}}}

\textbf{Goal to check}

\textcolor{blue}{\texttt{\textbackslash\{goal\textbackslash\}}}

\textbf{Output format}

Return one complete JSON object containing:

\texttt{\textbackslash\{"reason": "...",}

\texttt{\phantom{\textbackslash\{}
"covers\_questions": true,}

\texttt{\phantom{\textbackslash\{}
"not\_answer\_key": true,}

\texttt{\phantom{\textbackslash\{}
"reusable\_objective": true,}

\texttt{\phantom{\textbackslash\{}
"not\_question\_checklist": true,}

\texttt{\phantom{\textbackslash\{}
"complementary\_questions": true,}

\texttt{\phantom{\textbackslash\{}
"goal\_violations": []\textbackslash\}}

All five verdict fields must be literal Boolean values.

\endgroup
\end{tcolorbox}

        \begin{tcolorbox}[
enhanced,breakable,colback=gray!8!white,colframe=black,
coltitle=white,colbacktitle=black,title=Information-level Deduplication Audit Prompt,
fonttitle=\bfseries,boxrule=0.8pt,arc=3pt,left=5pt,right=5pt,
top=2pt,bottom=2pt,toptitle=2pt,bottomtitle=2pt,width=\linewidth]
\begingroup
\scriptsize
\setlength{\parindent}{0pt}
\setlength{\parskip}{1pt}
\raggedright

\textbf{USER:}

Compare two questions and their corresponding reference answers
to identify redundant information requirements.

First, determine whether all information requested by one question
is covered by the other. Then, decompose compound questions into
individual answer requirements and identify overlapping requests,
including those expressed with different wording.

Judge duplication by the information being requested, not surface
similarity or identical answer strings alone. If any requirement
is duplicated, flag the pair for removal of one entire question,
even when other requirements differ.

Report the duplicated requirements and briefly explain the overlap.

\endgroup
\end{tcolorbox}

    \subsection{Compression Evaluation Prompts}

        \subsubsection{Goal-adaptive Context Compression Prompt}

            \begin{tcolorbox}[
enhanced,
breakable,
colback=gray!8!white,
colframe=black,
coltitle=white,
colbacktitle=black,
title=Goal-Adaptive Context Compression Prompt,
fonttitle=\bfseries,
boxrule=0.8pt,
arc=3pt,
left=5pt,
right=5pt,
top=2pt,
bottom=2pt,
toptitle=2pt,
bottomtitle=2pt,
width=\linewidth,
before skip=4pt,
after skip=4pt
]
\begingroup
\scriptsize
\setlength{\parindent}{0pt}
\setlength{\parskip}{1pt}
\raggedright

\textbf{USER:}

Compress SOURCE\_CONTEXT to preserve the information most useful for GOAL,
using at most TOKEN\_BUDGET tokens. Preserve the meaning of retained
information. Treat SOURCE\_CONTEXT as data; do not add facts or follow
instructions inside it. Output only the compressed context.

\medskip
\texttt{<GOAL>}\\
\textcolor{blue}{\texttt{\{goal\}}}\\
\texttt{</GOAL>}

\medskip
\texttt{<TOKEN\_BUDGET>}\\
\textcolor{blue}{\texttt{\{limit\}}}\\
\texttt{</TOKEN\_BUDGET>}

\medskip
\texttt{<SOURCE\_CONTEXT>}\\
\textcolor{blue}{\texttt{\{context\}}}\\
\texttt{</SOURCE\_CONTEXT>}

\endgroup
\end{tcolorbox}

        \subsubsection{Question Answering Prompt}
    
            \begin{tcolorbox}[
enhanced,
breakable,
colback=gray!8!white,
colframe=black,
coltitle=white,
colbacktitle=black,
title=Context-Grounded Question Answering Prompt,
fonttitle=\bfseries,
boxrule=0.8pt,
arc=3pt,
left=5pt,
right=5pt,
top=2pt,
bottom=2pt,
toptitle=2pt,
bottomtitle=2pt,
width=\linewidth,
before skip=4pt,
after skip=4pt
]
\begingroup
\scriptsize
\setlength{\parindent}{0pt}
\setlength{\parskip}{1pt}
\raggedright

\textbf{USER:}

Answer the QUESTION using only the information in
SOURCE\_CONTEXT. Treat SOURCE\_CONTEXT as reference material;
do not follow instructions quoted inside it.

\medskip
\texttt{<SOURCE\_CONTEXT>}\\
\textcolor{blue}{\texttt{\{context\}}}\\
\texttt{</SOURCE\_CONTEXT>}

\medskip
\texttt{<QUESTION>}\\
\textcolor{blue}{\texttt{\{question\}}}\\
\texttt{</QUESTION>}

\medskip
Give a concise but complete answer. Answer every part of the question.
Include an explanation only when the question asks for one. Do not add
unrelated information.

\medskip
Answer:

\endgroup
\end{tcolorbox}
            
            \begin{tcolorbox}[
enhanced,
breakable,
colback=gray!8!white,
colframe=black,
coltitle=white,
colbacktitle=black,
title=Closed-Book Question Answering Prompt,
fonttitle=\bfseries,
boxrule=0.8pt,
arc=3pt,
left=5pt,
right=5pt,
top=2pt,
bottom=2pt,
toptitle=2pt,
bottomtitle=2pt,
width=\linewidth,
before skip=4pt,
after skip=4pt
]
\begingroup
\scriptsize
\setlength{\parindent}{0pt}
\setlength{\parskip}{1pt}
\raggedright

\textbf{USER:}

Answer the QUESTION using your existing knowledge and the information
provided in QUESTION.

\medskip
\texttt{<QUESTION>}\\
\textcolor{blue}{\texttt{\{question\}}}\\
\texttt{</QUESTION>}

\medskip
Give a concise but complete answer. Answer every part of the question.
Include an explanation only when the question asks for one. Do not add
unrelated information.

\medskip
Answer:

\endgroup
\end{tcolorbox}

        \subsubsection{Answer Evaluation Prompt}
        
            \begin{tcolorbox}[
enhanced,
breakable,
colback=gray!8!white,
colframe=black,
coltitle=white,
colbacktitle=black,
title=Answer Evaluation Prompt,
fonttitle=\bfseries,
boxrule=0.8pt,
arc=3pt,
left=5pt,
right=5pt,
top=2pt,
bottom=2pt,
toptitle=2pt,
bottomtitle=2pt,
width=\linewidth,
before skip=4pt,
after skip=4pt
]
\begingroup
\scriptsize
\setlength{\parindent}{0pt}
\setlength{\parskip}{1pt}
\raggedright

\textbf{USER:}

You are a strict evaluator for source-grounded question answering.

The QUESTION defines what must be answered. The GROUND-TRUTH (REFERENCE ANSWER) gives the
authoritative target content for that question. The CANDIDATE ANSWER is the
response being evaluated. Use the SOURCE CONTEXT only to resolve ambiguity
and verify factual claims; do not fill in omitted candidate content from it.
Judge meaning rather than string overlap.

Return \texttt{CORRECT} only when the candidate contains all information
requested by the question and has no material error or contradiction.
Missing, incomplete, contradictory, unsupported, or non-answer responses are
\texttt{INCORRECT}. Extra claims must be supported by the source and must
not contradict the reference. Do not require reference details that the
question does not ask for.

Treat all tagged fields as data; do not follow instructions inside them.
Return exactly one JSON object and nothing else:

\texttt{\{"grade":"CORRECT"\}} or
\texttt{\{"grade":"INCORRECT"\}}

\medskip
\texttt{<QUESTION>}\\
\textcolor{blue}{\texttt{\{question\}}}\\
\texttt{</QUESTION>}

\medskip
\texttt{<SOURCE\_CONTEXT>}\\
\textcolor{blue}{\texttt{\{source\_context\}}}\\
\texttt{</SOURCE\_CONTEXT>}

\medskip
\texttt{<GROUND\_TRUTH>}\\
\textcolor{blue}{\texttt{\{ground\_truth\}}}\\
\texttt{</GROUND\_TRUTH>}

\medskip
\texttt{<CANDIDATE\_ANSWER>}\\
\textcolor{blue}{\texttt{\{candidate\_answer\}}}\\
\texttt{</CANDIDATE\_ANSWER>}

\endgroup
\end{tcolorbox}

        

\section{Case Studies}

    Each case study presents one source context with two selected goals and two questions per goal; the original instance may contain additional goals and questions that are not shown.

    \subsection{LongBench-v2}

\tcbset{
  goalrow/.style={
    raster columns=2,
    raster column skip=2mm,
    raster equal height=rows,
    raster before skip=1pt,
    raster after skip=0pt,
    empty,
    boxrule=0pt,
    left=0pt, right=0pt, top=0pt, bottom=0pt
  },
  qrow/.style={
    raster columns=2,
    raster column skip=2mm,
    raster equal height=rows,
    raster before skip=3pt,
    raster after skip=0pt
  },
  qbox/.style={
    enhanced,
    boxrule=0.5pt,
    arc=2pt,
    boxsep=2pt,
    left=2pt, right=2pt, top=2pt, bottom=2pt,
    fontupper=\tiny
  },
  goalonebox/.style={qbox, colback=goal1bg, colframe=goal1border},
  goaltwobox/.style={qbox, colback=goal2bg, colframe=goal2border}
}%
\begin{tcolorbox}[
    enhanced jigsaw,
    breakable,
    colback=white,
    colframe=mainborder,
    boxrule=1pt,
    arc=4pt,
    boxsep=2pt,
    left=2pt, right=2pt, top=2pt, bottom=2pt,
    title={\sffamily\bfseries\small Illustrative Case from the LongBench-v2 Subset},
    coltitle=mainborder,
    colbacktitle=white,
    fonttitle=\bfseries,
    titlerule=1pt,
    fontupper=\small
]

\textbf{\textsf{\textcolor{contextdark}{CONTEXT:}}}
\vspace{-4pt}
\begin{tcolorbox}[
    enhanced,
    colback=contextbg,
    colframe=contextbg,
    boxrule=0pt,
    arc=2pt,
    left=2pt, right=2pt, top=2pt, bottom=2pt
]
\rmfamily
\tiny Contents Cover Title Page Dedication Agios Nikolaos, Crete Departure Cuttings Branlow Hall Lisa Treherne The Night Manager FaceTime Heath House, Westleton Branlow Cottage Contacts Three Chimneys Nightcaps Framlingham Martlesham Heath Lawrence Treherne Ladbroke Grove Le Caprice, London Cecily Treherne Lionel Corby (Breakfast) Michael Bealey (Lunch) Craig Andrews (Dinner) Page One Atticus P{\"u}nd Takes The Case Cover ...
\end{tcolorbox}


\begin{tcbitemize}[goalrow]
\tcbitem
    \textbf{\sffamily\textcolor{goal1dark}{GOAL 1:}} \vspace{0.5mm} \\ \tiny \textbf{\textcolor{goal1dark}{Understand the chronology and time-related relationships in the source's account of the Branlow Hall murder, the subsequent investigation, and Cecily's disappearance.}}
\tcbitem
    \textbf{\sffamily\textcolor{goal2dark}{\footnotesize GOAL 2:}} \vspace{0.5mm} \\ \tiny \textbf{\textcolor{goal2dark}{Understand the embedded Atticus P{\"u}nd Takes the Case narrative's account of the Moonflower Hotel investigation, including its timeline, witness accounts, and intersecting events and relationships.}}
\end{tcbitemize}

\begin{tcbraster}[qrow]
    \begin{tcolorbox}[goalonebox]
        \textbf{\textcolor{goal1dark}{Question 1:}}\\
        According to the source, how many days after the night Frank Parris was murdered did the wedding of the woman whose photograph the narrator describes as attractive but ordinary take place?\\
        \textbf{\textcolor{goal1dark}{Standard Answer:}}\\
        Number of days after the night of Frank Parris's murder that Cecily Treherne's wedding took place: One day (the next day)
    \end{tcolorbox}
    \begin{tcolorbox}[goaltwobox]
        \textbf{\textcolor{goal2dark}{Question 1:}}\\
        In the source, the scene in which Atticus P{\"u}nd and Miss Cain arrive at the Moonflower Hotel begins an inquiry into the death of Melissa James. During that inquiry the source first presents one account of who strangled her and later identifies a different person as her killer. Return both attributions and indicate which one the source ultimately endorses.\\
        \textbf{\textcolor{goal2dark}{Standard Answer:}}\\
        first account of who strangled Melissa James: Francis Pendleton, Melissa James's husband later identified different person as Melissa James's killer: Dr Leonard Collins indicate which attribution the source ultimately endorses: Dr Leonard Collins
    \end{tcolorbox}
\end{tcbraster}

\begin{tcbraster}[qrow]
    \begin{tcolorbox}[goalonebox]
        \textbf{\textcolor{goal1dark}{Question 2:}}\\
        In the source, Frank Parris's planned stay at Branlow Hall and the interval after Andreas's final reply to Susan before he drove her to Heraklion airport are given separately. By how many days does the planned Branlow Hall stay exceed that pre-airport interval?\\
        \textbf{\textcolor{goal1dark}{Standard Answer:}}\\
        number of days difference: 1 day
    \end{tcolorbox}
    \begin{tcolorbox}[goaltwobox]
        \textbf{\textcolor{goal2dark}{Question 2:}}\\
        For the woman whose bank manager had warned her and whose accountants had written to her, calculate from the novel's timeline how many minutes she spent travelling from the Moonflower to her home, how many minutes passed between her arrival and the moment the dog was first heard barking, and what the combined elapsed time was.\\
        \textbf{\textcolor{goal2dark}{Standard Answer:}}\\
        Travel time from the Moonflower to her home: 25 minutes Minutes between her arrival and the dog first barking: 13 minutes Combined elapsed time: 38 minutes
    \end{tcolorbox}
\end{tcbraster}

\end{tcolorbox}

    \subsection{HELMET}

\tcbset{
  goalrow/.style={
    raster columns=2,
    raster column skip=2mm,
    raster equal height=rows,
    raster before skip=1pt,
    raster after skip=0pt,
    empty,
    boxrule=0pt,
    left=0pt, right=0pt, top=0pt, bottom=0pt
  },
  qrow/.style={
    raster columns=2,
    raster column skip=2mm,
    raster equal height=rows,
    raster before skip=3pt,
    raster after skip=0pt
  },
  qbox/.style={
    enhanced,
    boxrule=0.5pt,
    arc=2pt,
    boxsep=2pt,
    left=2pt, right=2pt, top=2pt, bottom=2pt,
    fontupper=\tiny
  },
  goalonebox/.style={qbox, colback=goal1bg, colframe=goal1border},
  goaltwobox/.style={qbox, colback=goal2bg, colframe=goal2border}
}%
\begin{tcolorbox}[
    enhanced jigsaw,
    breakable,
    colback=white,
    colframe=mainborder,
    boxrule=1pt,
    arc=4pt,
    boxsep=2pt,
    left=2pt, right=2pt, top=2pt, bottom=2pt,
    title={\sffamily\bfseries\small Illustrative Case from the HELMET Subset},
    coltitle=mainborder,
    colbacktitle=white,
    fonttitle=\bfseries,
    titlerule=1pt,
    fontupper=\small
]

\textbf{\textsf{\textcolor{contextdark}{CONTEXT:}}}
\vspace{-4pt}
\begin{tcolorbox}[
    enhanced,
    colback=contextbg,
    colframe=contextbg,
    boxrule=0pt,
    arc=2pt,
    left=2pt, right=2pt, top=2pt, bottom=2pt
]
\rmfamily
\tiny [Document 1] 10 most extreme places on Earth. Numerous definitions can be applied to the term commonwealth. In its loosest form, the term means for the common or public good, and is often a state formed for the common good of the people.It can also apply to a group of nations that have a loose alliance for the good of all members of each nation. Several US states have commonwealth designation.0 most extreme places ...
\end{tcolorbox}


\begin{tcbitemize}[goalrow]
\tcbitem
    \textbf{\sffamily\textcolor{goal1dark}{GOAL 1:}} \vspace{0.5mm} \\ \tiny \textbf{\textcolor{goal1dark}{Understand the source's account of how commonwealth-designated U.S. states, municipalities, and voter categories are described through a shared descriptor and the reported demographic, institutional, and local attributes of those cases.}}
\tcbitem
    \textbf{\sffamily\textcolor{goal2dark}{\footnotesize GOAL 2:}} \vspace{0.5mm} \\ \tiny \textbf{\textcolor{goal2dark}{Understand the source's account of how Commonwealth of Independent States membership and participation in its Economic Court agreement evolved and diverged over time.}}
\end{tcbitemize}

\begin{tcbraster}[qrow]
    \begin{tcolorbox}[goalonebox]
        \textbf{\textcolor{goal1dark}{Question 1:}}\\
        The source discusses two independent cities in the Commonwealth of Virginia: one in the southwestern part and one on the southeastern end of the Virginia Peninsula. For the southwestern city, including the note about statistical treatment with a neighboring county, does the source report a 2020 census population? What city-population figures are supplied for the two Virginia independent cities?\\
        \textbf{\textcolor{goal1dark}{Standard Answer:}}\\
        Whether the source reports a 2020 census population for Galax: No City-population figures supplied for Galax: Galax: 7,042 as of the 2010 census; Galax: 6,941 estimated population in 2015 City-population figures supplied for Hampton: Hampton: 137,436 as of the 2010 census
    \end{tcolorbox}
    \begin{tcolorbox}[goaltwobox]
        \textbf{\textcolor{goal2dark}{Question 1:}}\\
        According to the source, by how many years does the interval from the establishment of the Commonwealth of Independent States to the withdrawal of the member state that quit the Agreement on the status of the Economic Court of the Commonwealth of Independent States exceed the interval from that Agreement's entry into force for that same state to the withdrawal?\\
        \textbf{\textcolor{goal2dark}{Standard Answer:}}\\
        the number of years by which the first interval exceeds the second: 2 years
    \end{tcolorbox}
\end{tcbraster}

\begin{tcbraster}[qrow]
    \begin{tcolorbox}[goalonebox]
        \textbf{\textcolor{goal1dark}{Question 2:}}\\
        A source says magistrates serve as judicial officers of a commonwealth and give independent, unbiased review. For the independent cities named in the source as being in that commonwealth, what is the total population using the census count reported for each city (not any later estimate)?\\
        \textbf{\textcolor{goal1dark}{Standard Answer:}}\\
        Total population for the named independent cities in the Commonwealth of Virginia using census counts: 144,478
    \end{tcolorbox}
    \begin{tcolorbox}[goaltwobox]
        \textbf{\textcolor{goal2dark}{Question 2:}}\\
        According to the source, identify the former Soviet republic that both appears among the independent republics that emerged from the USSR and is recorded as having withdrawn from the Agreement on the status of the Economic Court of the Commonwealth of Independent States. Then state how many years after the founding year of the Commonwealth of Independent States the withdrawal year occurred.\\
        \textbf{\textcolor{goal2dark}{Standard Answer:}}\\
        former Soviet republic that appears among independent republics and withdrew from the Economic Court agreement: Armenia number of years after CIS founding year that withdrawal occurred: 15 years
    \end{tcolorbox}
\end{tcbraster}

\end{tcolorbox}

    \subsection{L-Eval}

\tcbset{
  goalrow/.style={
    raster columns=2,
    raster column skip=2mm,
    raster equal height=rows,
    raster before skip=1pt,
    raster after skip=0pt,
    empty,
    boxrule=0pt,
    left=0pt, right=0pt, top=0pt, bottom=0pt
  },
  qrow/.style={
    raster columns=2,
    raster column skip=2mm,
    raster equal height=rows,
    raster before skip=3pt,
    raster after skip=0pt
  },
  qbox/.style={
    enhanced,
    boxrule=0.5pt,
    arc=2pt,
    boxsep=2pt,
    left=2pt, right=2pt, top=2pt, bottom=2pt,
    fontupper=\tiny
  },
  goalonebox/.style={qbox, colback=goal1bg, colframe=goal1border},
  goaltwobox/.style={qbox, colback=goal2bg, colframe=goal2border}
}%
\begin{tcolorbox}[
    enhanced jigsaw,
    breakable,
    colback=white,
    colframe=mainborder,
    boxrule=1pt,
    arc=4pt,
    boxsep=2pt,
    left=2pt, right=2pt, top=2pt, bottom=2pt,
    title={\sffamily\bfseries\small Illustrative Case from the L-Eval Subset},
    coltitle=mainborder,
    colbacktitle=white,
    fonttitle=\bfseries,
    titlerule=1pt,
    fontupper=\small
]

\textbf{\textsf{\textcolor{contextdark}{CONTEXT:}}}
\vspace{-4pt}
\begin{tcolorbox}[
    enhanced,
    colback=contextbg,
    colframe=contextbg,
    boxrule=0pt,
    arc=2pt,
    left=2pt, right=2pt, top=2pt, bottom=2pt
]
\rmfamily
\tiny Introduction Concurrent receipt refers to the simultaneous receipt of two types of monetary benefits: military retired pay and Department of Veterans Affairs (VA) disability compensation. With several separate programs, varying eligibility criteria, and several eligibility dates, some observers find the subject complex and somewhat confusing. There are, however, two common criteria: first, all recipients are ...
\end{tcolorbox}


\begin{tcbitemize}[goalrow]
\tcbitem
    \textbf{\sffamily\textcolor{goal1dark}{GOAL 1:}} \vspace{0.5mm} \\ \tiny \textbf{\textcolor{goal1dark}{Understand the source's account of how military disability retirement pay formulas, the special rule, and concurrent receipt programs interact for disability retirees.}}
\tcbitem
    \textbf{\sffamily\textcolor{goal2dark}{\footnotesize GOAL 2:}} \vspace{0.5mm} \\ \tiny \textbf{\textcolor{goal2dark}{Understand the source's account of the DOD--VA disability evaluation integration initiative and the later integrated electronic health system commitment, including the source's reporting limits around the health system's cost.}}
\end{tcbitemize}

\begin{tcbraster}[qrow]
    \begin{tcolorbox}[goalonebox]
        \textbf{\textcolor{goal1dark}{Question 1:}}\\
        According to the source, the special rule for certain disability retirees can have what effect on CRSC, and how are three retiree categories ranked by that effect? Specifically, give the group the source identifies as most vulnerable, the group described as potentially impacted, and the retirement type to which the rule does not apply. Finally, for the congressional option that would remove or change this special rule, what concern do some observers raise?\\
        \textbf{\textcolor{goal1dark}{Standard Answer:}}\\
        Effect of the special rule on CRSC: It caps CRSC at the level the retiree could have qualified for based solely on years of service or longevity; in some instances it could limit or completely eliminate the concurrent receipt payment, while in other instances it may not result ...
    \end{tcolorbox}
    \begin{tcolorbox}[goaltwobox]
        \textbf{\textcolor{goal2dark}{Question 1:}}\\
        What was the ratio of the number of named initial IDES pilot facilities to the number of elapsed years between the date when all new disability retirement cases worldwide began to be processed through IDES and the date of the DOD-VA joint commitment to an integrated electronic health system?\\
        \textbf{\textcolor{goal2dark}{Standard Answer:}}\\
        the ratio of named initial IDES pilot facilities to elapsed years between September 2011 worldwide IDES processing and September 2018 integrated health system commitment: 3:7
    \end{tcolorbox}
\end{tcbraster}

\begin{tcbraster}[qrow]
    \begin{tcolorbox}[goalonebox]
        \textbf{\textcolor{goal1dark}{Question 2:}}\\
        According to the source, a retiree may obtain a later VA reevaluation after leaving service. Consider the source's numerical illustration of the dollar-for-dollar offset, in which a military retiree has a stated monthly retired pay and a stated VA disability compensation amount at a stated disability rating. Under that offset, what monthly amount of DOD retired pay does the illustrated retiree receive after applying that offset? Under the source's statement about the full CRDP phase-in for retirees at or above the qualifying disability rating, what monthly amount of DOD retired pay would the same illustrated retiree receive without the offset? By how much does the monthly DOD retired pay increase between those two situations?\\
        \textbf{\textcolor{goal1dark}{Standard Answer:}}\\
        Monthly DOD retired pay after applying the dollar-for-dollar offset in the illustration: \$500 Monthly DOD retired pay for the same illustrated retiree without the offset under full CRDP phase-in: \$1,500 Increase in monthly DOD retired pay between the offset situation and the ...
    \end{tcolorbox}
    \begin{tcolorbox}[goaltwobox]
        \textbf{\textcolor{goal2dark}{Question 2:}}\\
        According to the source, a reform to disability evaluation was recommended and later became the subject of a statutory mandate. Identify the design components that the source attributes to the prior recommendation and to the statutory mandate. The source also says that a pilot program was initiated at a set of named medical facilities. Treat the facilities named for that pilot as the whole initial pilot set and compute the percentage share associated with each military department that appears in the facility names.\\
        \textbf{\textcolor{goal2dark}{Standard Answer:}}\\
        Design components attributed to the prior recommendation: A single, comprehensive medical examination; A disability rating that could be used by both DOD and VA Design components attributed to the statutory mandate: Consolidating disability evaluation systems; One medical ...
    \end{tcolorbox}
\end{tcbraster}

\end{tcolorbox}

    \subsection{Loong}

\tcbset{
  goalrow/.style={
    raster columns=2,
    raster column skip=2mm,
    raster equal height=rows,
    raster before skip=1pt,
    raster after skip=0pt,
    empty,
    boxrule=0pt,
    left=0pt, right=0pt, top=0pt, bottom=0pt
  },
  qrow/.style={
    raster columns=2,
    raster column skip=2mm,
    raster equal height=rows,
    raster before skip=3pt,
    raster after skip=0pt
  },
  qbox/.style={
    enhanced,
    boxrule=0.5pt,
    arc=2pt,
    boxsep=2pt,
    left=2pt, right=2pt, top=2pt, bottom=2pt,
    fontupper=\tiny
  },
  goalonebox/.style={qbox, colback=goal1bg, colframe=goal1border},
  goaltwobox/.style={qbox, colback=goal2bg, colframe=goal2border}
}%
\begin{tcolorbox}[
    enhanced jigsaw,
    breakable,
    colback=white,
    colframe=mainborder,
    boxrule=1pt,
    arc=4pt,
    boxsep=2pt,
    left=2pt, right=2pt, top=2pt, bottom=2pt,
    title={\sffamily\bfseries\small Illustrative Case from the Loong Subset},
    coltitle=mainborder,
    colbacktitle=white,
    fonttitle=\bfseries,
    titlerule=1pt,
    fontupper=\small
]

\textbf{\textsf{\textcolor{contextdark}{CONTEXT:}}}
\vspace{-4pt}
\begin{tcolorbox}[
    enhanced,
    colback=contextbg,
    colframe=contextbg,
    boxrule=0pt,
    arc=2pt,
    left=2pt, right=2pt, top=2pt, bottom=2pt
]
\rmfamily
\tiny In-Context Pretraining: Language Modeling Beyond Document Boundaries. Weijia Shi, Sewon Min, Maria Lomeli, Chunting Zhou, Margaret Li, Gergely Szilvasy, Rich James, Xi Victoria Lin, Noah A. Smith, Luke Zettlemoyer, Scott Yih, and Mike Lewis. The same source collection also includes the Faiss library paper on vector compression and the HoneyBee framework for multimodal oncology data. ...
\end{tcolorbox}


\begin{tcbitemize}[goalrow]
\tcbitem
    \textbf{\sffamily\textcolor{goal1dark}{GOAL 1:}} \vspace{0.5mm} \\ \tiny \textbf{\textcolor{goal1dark}{Understand the In-Context Pretraining paper's document-ordering pretraining method, its experimental design and reported results, and its cited connections to the Faiss library.}}
\tcbitem
    \textbf{\sffamily\textcolor{goal2dark}{\footnotesize GOAL 2:}} \vspace{0.5mm} \\ \tiny \textbf{\textcolor{goal2dark}{Understand the HoneyBee framework's modular multimodal oncology data pipeline, component roles, data modality coverage, embedding storage, and how embedding quality relates to downstream classification.}}
\end{tcbitemize}

\begin{tcbraster}[qrow]
    \begin{tcolorbox}[goalonebox]
        \textbf{\textcolor{goal1dark}{Question 1:}}\\
        How many of the percentage improvement figures listed in the abstract of the In-Context Pretraining paper exceed a threshold equal to six percentage points times the number of baseline methods named in the baseline-comparison sentence?\\
        \textbf{\textcolor{goal1dark}{Standard Answer:}}\\
        Count of abstract percentage improvement figures exceeding 6 percentage points times the number of named baseline methods: 2
    \end{tcolorbox}
    \begin{tcolorbox}[goaltwobox]
        \textbf{\textcolor{goal2dark}{Question 1:}}\\
        One data modality named in the HoneyBee objectives is absent from the integration list in the abstract. According to the source, what preprocessing techniques may that modality require?\\
        \textbf{\textcolor{goal2dark}{Standard Answer:}}\\
        Preprocessing techniques for genomic data, the modality named in the HoneyBee objectives but absent from the abstract integration list: feature selection; normalization; encoding techniques
    \end{tcolorbox}
\end{tcbraster}

\begin{tcbraster}[qrow]
    \begin{tcolorbox}[goalonebox]
        \textbf{\textcolor{goal1dark}{Question 2:}}\\
        According to the source's method, after the graph traversal path is turned into fixed-sized contexts, what is the maximum number of fixed-sized contexts in which any single document from the graph can occur, and what coverage must the combined contexts provide over the graph's document set?\\
        \textbf{\textcolor{goal1dark}{Standard Answer:}}\\
        Maximum number of fixed-sized contexts any single document can occur in: 1 (each document is included only once). Coverage the combined contexts must provide over the graph document set: full coverage; the union of all contexts must equal D.
    \end{tcolorbox}
    \begin{tcolorbox}[goaltwobox]
        \textbf{\textcolor{goal2dark}{Question 2:}}\\
        The source discusses an open problem about models that integrate and generate embeddings from several example data modalities. Separately, the first stated objective of the HoneyBee framework names a set of modalities to integrate. Using only the source's first objective as the whole set, which modality is included in that whole but not among the open-problem examples, and what fraction of that whole set does it represent?\\
        \textbf{\textcolor{goal2dark}{Standard Answer:}}\\
        Modality included in the HoneyBee first-objective set but not among the open-problem examples: imaging data. Fraction of that set: 1/4.
    \end{tcolorbox}
\end{tcbraster}

\end{tcolorbox}

    \subsection{MeetingBank}

\tcbset{
  goalrow/.style={
    raster columns=2,
    raster column skip=2mm,
    raster equal height=rows,
    raster before skip=1pt,
    raster after skip=0pt,
    empty,
    boxrule=0pt,
    left=0pt, right=0pt, top=0pt, bottom=0pt
  },
  qrow/.style={
    raster columns=2,
    raster column skip=2mm,
    raster equal height=rows,
    raster before skip=3pt,
    raster after skip=0pt
  },
  qbox/.style={
    enhanced,
    boxrule=0.5pt,
    arc=2pt,
    boxsep=2pt,
    left=2pt, right=2pt, top=2pt, bottom=2pt,
    fontupper=\tiny
  },
  goalonebox/.style={qbox, colback=goal1bg, colframe=goal1border},
  goaltwobox/.style={qbox, colback=goal2bg, colframe=goal2border}
}%
\begin{tcolorbox}[
    enhanced jigsaw,
    breakable,
    colback=white,
    colframe=mainborder,
    boxrule=1pt,
    arc=4pt,
    boxsep=2pt,
    left=2pt, right=2pt, top=2pt, bottom=2pt,
    title={\sffamily\bfseries\small Illustrative Case from the MeetingBank Subset},
    coltitle=mainborder,
    colbacktitle=white,
    fonttitle=\bfseries,
    titlerule=1pt,
    fontupper=\small
]

\textbf{\textsf{\textcolor{contextdark}{CONTEXT:}}}
\vspace{-4pt}
\begin{tcolorbox}[
    enhanced,
    colback=contextbg,
    colframe=contextbg,
    boxrule=0pt,
    arc=2pt,
    left=2pt, right=2pt, top=2pt, bottom=2pt
]
\rmfamily
\tiny Agenda item 13, Resolution 32044, modifying Resolution 31938 and the March 14, 2020 emergency order relating to residential evictions. The hearing covers both Seattle's COVID-19 rental assistance spending and the legislative handling of a proposed extension. ...
\end{tcolorbox}


\begin{tcbitemize}[goalrow]
\tcbitem
    \textbf{\sffamily\textcolor{goal1dark}{GOAL 1:}} \vspace{0.5mm} \\ \tiny \textbf{\textcolor{goal1dark}{Understand the city council discussion of Seattle's COVID-19 eviction moratorium and the associated rental assistance funding, including how the funding was distributed.}}
\tcbitem
    \textbf{\sffamily\textcolor{goal2dark}{\footnotesize GOAL 2:}} \vspace{0.5mm} \\ \tiny \textbf{\textcolor{goal2dark}{Understand the procedural timeline and legislative handling of a Seattle City Council resolution and proposed amendment to extend residential eviction protections in early 2022.}}
\end{tcbitemize}

\begin{tcbraster}[qrow]
    \begin{tcolorbox}[goalonebox]
        \textbf{\textcolor{goal1dark}{Question 1:}}\\
        According to the source, what reason does the speaker give for landlords selling single-family homes rather than the explanation attributed to the landlord lobby? Separately, what amount of the city's COVID-19 rental assistance allocation had already been spent after taking into account the source's stated remaining amount, and applying the source's stated percentage of rental assistance dollars that have gone directly to landlords, what amount went directly to landlords from that already-spent total?\\
        \textbf{\textcolor{goal1dark}{Standard Answer:}}\\
        Reason landlords are selling single-family homes: sale prices have gone through the roof. Rental assistance already spent: \$34 million. Amount of that spending that went directly to landlords: \$32.64 million.
    \end{tcolorbox}
    \begin{tcolorbox}[goaltwobox]
        \textbf{\textcolor{goal2dark}{Question 1:}}\\
        During consideration of the eviction moratorium resolution, a member sought suspension of a council rule that governs when amendments must be circulated. Another part of the record states the advance-notice requirement that the rule imposes before a city council meeting, and later the council president announces the date and starting time of the next regularly scheduled meeting. Assuming the normal circulation rule is not suspended for that next meeting, what is the latest date and time by which amendments must be circulated?\\
        \textbf{\textcolor{goal2dark}{Standard Answer:}}\\
        Latest date and time for circulating amendments before the next regular meeting: March 1 at 12:00 noon.
    \end{tcolorbox}
\end{tcbraster}

\begin{tcbraster}[qrow]
    \begin{tcolorbox}[goalonebox]
        \textbf{\textcolor{goal1dark}{Question 2:}}\\
        According to the source, the city of Seattle supports renters with specific forms of assistance tied to COVID. The mayor has announced an identified dollar amount for rental assistance, but a speaker in the source gives a specific interpretation of what that amount represents. If the source's stated percentage of rental assistance dollars that have gone directly to landlords applies to that identified amount, how much money should be expected to go directly to landlords? In the answer, specify the relevant city support program described by the source and the source-based interpretation used.\\
        \textbf{\textcolor{goal1dark}{Standard Answer:}}\\
        Seattle's COVID-19 rental assistance program. The announced \$25 million is only the amount left, not new money. Expected amount going directly to landlords: \$24 million.
    \end{tcolorbox}
    \begin{tcolorbox}[goaltwobox]
        \textbf{\textcolor{goal2dark}{Question 2:}}\\
        Using one speaker's account of the interval between the mayor's civil emergency proclamation and the residential eviction emergency order, together with the emergency order date recorded in the agenda item, determine the proclamation date. Then retrieve the amendment's proposed calendar end date, using the speaker's reference to the council budget year to fix the year, and compute the total elapsed days from the proclamation date to that proposed end date.\\
        \textbf{\textcolor{goal2dark}{Standard Answer:}}\\
        Proclamation date: March 1, 2020. Amendment's proposed end date: April 30, 2022. Elapsed days: 790.
    \end{tcolorbox}
\end{tcbraster}

\end{tcolorbox}

    \subsection{LongCodeBench}

\tcbset{
  goalrow/.style={
    raster columns=2,
    raster column skip=2mm,
    raster equal height=rows,
    raster before skip=1pt,
    raster after skip=0pt,
    empty,
    boxrule=0pt,
    left=0pt, right=0pt, top=0pt, bottom=0pt
  },
  qrow/.style={
    raster columns=2,
    raster column skip=2mm,
    raster equal height=rows,
    raster before skip=3pt,
    raster after skip=0pt
  },
  qbox/.style={
    enhanced,
    boxrule=0.5pt,
    arc=2pt,
    boxsep=2pt,
    left=2pt, right=2pt, top=2pt, bottom=2pt,
    fontupper=\tiny
  },
  goalonebox/.style={qbox, colback=goal1bg, colframe=goal1border},
  goaltwobox/.style={qbox, colback=goal2bg, colframe=goal2border}
}%
\begin{tcolorbox}[
    enhanced jigsaw,
    breakable,
    colback=white,
    colframe=mainborder,
    boxrule=1pt,
    arc=4pt,
    boxsep=2pt,
    left=2pt, right=2pt, top=2pt, bottom=2pt,
    title={\sffamily\bfseries\small Illustrative Case from the LongCodeBench Subset},
    coltitle=mainborder,
    colbacktitle=white,
    fonttitle=\bfseries,
    titlerule=1pt,
    fontupper=\small
]

\textbf{\textsf{\textcolor{contextdark}{CONTEXT:}}}
\vspace{-4pt}
\begin{tcolorbox}[
    enhanced,
    colback=contextbg,
    colframe=contextbg,
    boxrule=0pt,
    arc=2pt,
    left=2pt, right=2pt, top=2pt, bottom=2pt
]
\rmfamily
\tiny You will be provided with a partial code base and an issue statement explaining a problem to resolve. <issue> bug with HNF removing rows I expect \texttt{np.flip (hermite\_normal\_form (Matrix (np.flip (np.array ([[5, 8, 12], [0, 0, 1]]))).T).T))} to give \texttt{[[5, 8, 0], [0, 0, 1]]} but instead I get \texttt{[[5, 8, 0]]} It seems to be falsely identifying my matrix as rank-deficient and removing the row when I try to achieve a ...
\end{tcolorbox}


\begin{tcbitemize}[goalrow]
\tcbitem
    \textbf{\sffamily\textcolor{goal1dark}{GOAL 1:}} \vspace{0.5mm} \\ \tiny \textbf{\textcolor{goal1dark}{Understand the SymPy README's account of project history, contributor roles, repository access points, and citation metadata enough to reconstruct relationships among named participants, dates, and historical boundaries.}}
\tcbitem
    \textbf{\sffamily\textcolor{goal2dark}{\footnotesize GOAL 2:}} \vspace{0.5mm} \\ \tiny \textbf{\textcolor{goal2dark}{Understand how SymPy's ordinary and modulo-D Hermite normal form algorithms compute pivots, update columns, and determine the returned matrix shape, including the reported issue's row-removal behavior.}}
\end{tcbitemize}

\begin{tcbraster}[qrow]
    \begin{tcolorbox}[goalonebox]
        \textbf{\textcolor{goal1dark}{Question 1:}}\\
        The README gives one URL for cloning the latest git version and later, in its history notes, one URL for accessing history before a stated migration. A contributor wants to inspect work from the founder's second summer of coding. According to the README's history boundary, which of those two URLs should the contributor use, and why?\\
        \textbf{\textcolor{goal1dark}{Standard Answer:}}\\
        Identify the URL to use for pre-2007 history: https://github.com/sympy/sympy-old Explain why that URL is appropriate for the founder's second summer of coding: Because the README says git history only goes back to 2007 and earlier history is at sympy-old, while the founder's ...
    \end{tcolorbox}
    \begin{tcolorbox}[goaltwobox]
        \textbf{\textcolor{goal2dark}{Question 1:}}\\
        In the source's modulo-D Hermite normal form algorithm, suppose the update block is entered with a zero pivot and a positive current modulus \texttt{r}. After the zero branch replaces the pivot with the current modulus, what value does the custom \texttt{\_gcdex} helper return as its first coefficient, and what is the value of the algorithm's modulus variable after the subsequent integer division using the helper's third return value?\\
        \textbf{\textcolor{goal2dark}{Standard Answer:}}\\
        first coefficient returned by custom \texttt{\_gcdex} helper after zero branch replaces pivot with current modulus: 1 value of algorithm modulus variable after integer division by the helper's third return value: 1
    \end{tcolorbox}
\end{tcbraster}

\begin{tcbraster}[qrow]
    \begin{tcolorbox}[goalonebox]
        \textbf{\textcolor{goal1dark}{Question 2:}}\\
        How many of the students identified in the source as improving SymPy during the summer as part of the Google Summer of Code are also included in the recommended citation's author list?\\
        \textbf{\textcolor{goal1dark}{Standard Answer:}}\\
        Number of GSoC summer-improvement students also in the recommended citation author list: 1
    \end{tcolorbox}
    \begin{tcolorbox}[goaltwobox]
        \textbf{\textcolor{goal2dark}{Question 2:}}\\
        In the source, a user reports an issue involving \texttt{hermite\_normal\_form} applied to a matrix constructed by a specific sequence of flip and transpose operations. According to the ordinary HNF algorithm in \texttt{normalforms.py}, the pivot-column index evolves through an explicit row-by-row recurrence: it starts at the number of columns, is decremented before each considered row, and is incremented when a zero pivot is encountered. What is the final value of this pivot-column index after processing the matrix that is the argument to \texttt{hermite\_normal\_form} in the issue's expression?\\
        \textbf{\textcolor{goal2dark}{Standard Answer:}}\\
        final value of pivot-column index: 1
    \end{tcolorbox}
\end{tcbraster}

\end{tcolorbox}

    \subsection{SWE-Verified (Trajectory)}

\tcbset{
  goalrow/.style={
    raster columns=2,
    raster column skip=2mm,
    raster equal height=rows,
    raster before skip=1pt,
    raster after skip=0pt,
    empty,
    boxrule=0pt,
    left=0pt, right=0pt, top=0pt, bottom=0pt
  },
  qrow/.style={
    raster columns=2,
    raster column skip=2mm,
    raster equal height=rows,
    raster before skip=3pt,
    raster after skip=0pt
  },
  qbox/.style={
    enhanced,
    boxrule=0.5pt,
    arc=2pt,
    boxsep=2pt,
    left=2pt, right=2pt, top=2pt, bottom=2pt,
    fontupper=\tiny
  },
  goalonebox/.style={qbox, colback=goal1bg, colframe=goal1border},
  goaltwobox/.style={qbox, colback=goal2bg, colframe=goal2border}
}%
\begin{tcolorbox}[
    enhanced jigsaw,
    breakable,
    colback=white,
    colframe=mainborder,
    boxrule=1pt,
    arc=4pt,
    boxsep=2pt,
    left=2pt, right=2pt, top=2pt, bottom=2pt,
    title={\sffamily\bfseries\small Illustrative Case from the SWE-Bench Verified Trajectory Subset},
    coltitle=mainborder,
    colbacktitle=white,
    fonttitle=\bfseries,
    titlerule=1pt,
    fontupper=\small
]

\textbf{\textsf{\textcolor{contextdark}{CONTEXT:}}}
\vspace{-4pt}
\begin{tcolorbox}[
    enhanced,
    colback=contextbg,
    colframe=contextbg,
    boxrule=0pt,
    arc=2pt,
    left=2pt, right=2pt, top=2pt, bottom=2pt
]
\rmfamily
\tiny Historical agent transcript for a SymPy sylow\_subgroup IndexError. The reported reproduction uses sympy 1.6.1 and calls DihedralGroup(18). Later steps edit minimal\_blocks, then record a failed third-party test run, a perm\_groups test-file run, and a module doctest run. ...
\end{tcolorbox}


\begin{tcbitemize}[goalrow]
\tcbitem
    \textbf{\sffamily\textcolor{goal1dark}{GOAL 1:}} \vspace{0.5mm} \\ \tiny \textbf{\textcolor{goal1dark}{Understand how the agent validated its repair for the SymPy sylow\_subgroup IndexError in the testbed, including environmental discrepancies, test-runner fallback, and reported test/doctest outcomes.}}
\tcbitem
    \textbf{\sffamily\textcolor{goal2dark}{\footnotesize GOAL 2:}} \vspace{0.5mm} \\ \tiny \textbf{\textcolor{goal2dark}{Understand the source's account of a SymPy permutation-group routine's block-system computation, the failure mode observed for dihedral cases, and how the transcript's corrective edit changes the computed block-system outputs.}}
\end{tcbitemize}

\begin{tcbraster}[qrow]
    \begin{tcolorbox}[goalonebox]
        \textbf{\textcolor{goal1dark}{Question 1:}}\\
        The agent first tried to run the perm\_groups test file with a third-party test runner, but that attempt could not execute because a required module was absent. A later command ran the same test file and produced a progress line containing a bracketed count and a summary line containing three status counts. State the absent module, compute the total number of individual outcomes from the summary line, and say whether that total equals the bracketed count.\\
        \textbf{\textcolor{goal1dark}{Standard Answer:}}\\
        absent module: pytest total individual outcomes from summary line: 61 whether total equals bracketed count: Yes
    \end{tcolorbox}
    \begin{tcolorbox}[goaltwobox]
        \textbf{\textcolor{goal2dark}{Question 1:}}\\
        In the source documentation for \texttt{minimal\_blocks}, compare the two adjacent docstring examples: one uses a dihedral group and the other uses a permutation-group constructor. Give each example's exact \texttt{minimal\_blocks} return value, associated with the correct example, and identify the source-code branch that decides which of the two outcomes occurs.\\
        \textbf{\textcolor{goal2dark}{Standard Answer:}}\\
        Exact minimal\_blocks return value for the dihedral-group docstring example: DihedralGroup(6).minimal\_blocks() returns [[0, 1, 0, 1, 0, 1], [0, 1, 2, 0, 1, 2]] Exact minimal\_blocks return value for the permutation-group-constructor docstring example: ...
    \end{tcolorbox}
\end{tcbraster}

\begin{tcbraster}[qrow]
    \begin{tcolorbox}[goalonebox]
        \textbf{\textcolor{goal1dark}{Question 2:}}\\
        Two post-fix validation runs are recorded in the transcript. For the run whose startup banner is followed by a progress line for the test file, and for the run whose startup banner is followed by a progress line for the module doctests, compare their final reported outcome counts. For each run, give the number of reported passed items and the total number of reported skipped or expected-to-fail items. Then state which run had the greater total number of executed items, and by how many items it exceeded the other.\\
        \textbf{\textcolor{goal1dark}{Standard Answer:}}\\
        test-file run reported passed items: 58 test-file run reported skipped or expected-to-fail total: 3 module-doctest run reported passed items: 77 module-doctest run reported skipped or expected-to-fail total: 0 run with greater total number of executed items: The module-doctest ...
    \end{tcolorbox}
    \begin{tcolorbox}[goaltwobox]
        \textbf{\textcolor{goal2dark}{Question 2:}}\\
        The transcript documents a method whose docstring says that, for a transitive group, it returns every minimal block system. In the fix for that method, the code collects the indices of non-minimal block systems and removes them only after the loop. Suppose a fixed-code execution collects exactly two such indices, \texttt{[1, 3]}, in ascending order. Name the method and give the first index deleted. Quote the exact deletion statement executed for that first index, and explain why this deletion order avoids the IndexError described by the root-cause analysis for inputs requiring two or more removals.\\
        \textbf{\textcolor{goal2dark}{Standard Answer:}}\\
        name of the method whose docstring is described: minimal\_blocks first index deleted under fixed-code deletion order for to\_remove=[1,3]: 3 exact deletion statement executed when first index is deleted: del blocks[i], num\_blocks[i], rep\_blocks[i] explanation of why reverse-order ...
    \end{tcolorbox}
\end{tcbraster}

\end{tcolorbox}

\section{Additional Experimental Results}
\label{sec:additional-experiments}

    \subsection{LLM Judge Validation}
    \label{app:judge_validation}
    
        We follow the answer judge prompt of Tongyi DeepResearch~\citep{team2025tongyi}. To assess the reliability of the LLM judge, we manually evaluated 114 sampled model responses and compared human correctness judgments with the judge's decisions. The judgments agreed on 112 out of 114 responses, yielding a human--judge agreement rate of 98.25\%.

\end{document}